\documentclass[journal]{IEEEtran}

\usepackage{amsmath,epsfig,dsfont}
\usepackage{subfigure}
\usepackage{algorithm}
\usepackage{algorithmic}
\usepackage{amssymb,bm}
\usepackage{amsfonts}
\usepackage{stfloats}
\usepackage{multirow}
\ifCLASSINFOpdf
\else
\fi
\begin{document}

\title{Evolutionary Dynamics of Information Diffusion over Social Networks}
%
%
%
\author{\authorblockN{Chunxiao~Jiang\authorrefmark{1}\authorrefmark{2}, Yan~Chen\authorrefmark{1}, and K. J. Ray~Liu\authorrefmark{1}} \\ 
      \small\authorblockA{\authorrefmark{1}Department of Electrical and Computer Engineering, University of Maryland, College Park, MD 20742, USA\\ 
          \authorrefmark{2}Department of Electronic Engineering, Tsinghua University, Beijing, 100084, P. R. China\\ 
        E-mail:\{jcx, yan, kjrliu\}@umd.edu}}
\maketitle

\begin{abstract}
Current social networks are of extremely large-scale generating tremendous information flows at every moment. How information diffuse over social networks has attracted much attention from both industry and academics. Most of the existing works on information diffusion analysis are based on machine learning methods focusing on social network structure analysis and empirical data mining. However, the dynamics of information diffusion, which are heavily influenced by network users' decisions, actions and their socio-economic interactions, is generally ignored by most of existing works. In this paper, we propose an evolutionary game theoretic framework to model the dynamic information diffusion process in social networks. Specifically, we derive the information diffusion dynamics in complete networks, uniform degree and non-uniform degree networks, with the highlight of two special networks, Erd\H{o}s-R\'enyi random network and the Barab\'asi-Albert scale-free network. We find that the dynamics of information diffusion over these three kinds of networks are scale-free and the same with each other when the network scale is sufficiently large. To verify our theoretical analysis, we perform simulations for the information diffusion over synthetic networks and real-world Facebook networks. Moreover, we also conduct experiment on Twitter hashtags dataset, which shows that the proposed game theoretic model can well fit and predict the information diffusion over real social networks.
\end{abstract}
%
\section{Introduction}

Nowadays, the concept of ``social networks'' has become ubiquitous in our daily life. Due to its diverse implication, researchers from different disciplines have been working on social networking from various perspectives \cite{Liu}. A social network is a social structure made of a set of social actors (e.g., individuals, organizations, websits or even equipments) and a set of the relationships and interactions between these actors \cite{sns}. It plays a fundamental role as a medium for the diffusion of information, ideas, and influence among its users. With the rapid development of the Internet and mobile technologies, today's social networks are of extremely large scale, e.g., Facebook has nearly one billion active users as of September 2012 \cite{facebookuser}. Meanwhile, the information size on the social networks are becoming even tremendous-scale. For instance, there were 175 million tweets sent from Twitter every day throughout 2012 \cite{2012twitter}.

On a social network, various kinds of new information are originated everyday, e.g., when a new model of cell phone is announced, some deal advertisements are released, some political statements from a party are declared, or rumors about actors and singers are reported. Such information can either disappear quickly or inspire heated discussions. If one wants to understand or even predict the final destiny of a piece of new information, it would be rather important to study the dynamics of information diffusion over the underlying social network, which can be affected by multiple factors, such as a user's own interest on the information, the users' social interactions and influences with each other, as well as the ``word-of-mouth'' effects. Under such a circumstance, the topic of how information diffuse over social networks draws great attentions from both industry and academia recently. On one hand, the study of information diffusion can help the enterprises/polititians to achieve efficient and effective advertisement/advocation. On the other hand, from the security point of view, the study of information diffusion can also help to prevent the detrimental information spreading, e.g. computer virus, rumors and inauthentic news.

\begin{figure}[!t]
  \centering
  \centerline{\epsfig{figure=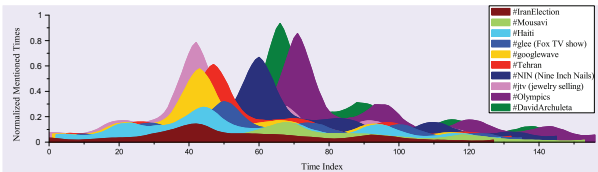,width=8cm}}
  \caption{Normalized number of forwarding of different Twitter hashtags using dataset \cite{dataset}, collected from Jun. to Dec. 2009.}\label{twitter}\vspace{-5mm}
\end{figure}

For example, Fig.\,\ref{twitter} shows the normalized mention times of different Twitter hashtags \cite{dataset}, which contains 1000 highest total volume hashtags among 6 million hashtags from Jun. to Dec. 2009. The highest peak is corresponding to the keyword ``DavidArchuleta'', the name of the singer who was the runner-up of \emph{American Idol} and won the Rising Male Star award on Sep. 2009. Such information dynamics of the singer in Fig.\,\ref{twitter} can be used to estimate the approval ratio, based on which his agency would adjust their further advertisement strategy. Another example is the Twitter political index \cite{twitterpi}, which provided the approval ratios of Obama and Romney during 2012 USA president election through mining Twitter users' online information diffusion and comment behaviors, especially after Obama or Romney delivered speeches. Based on the approval ratio, the candidates can adjust their attitudes in the next speeches to attract more supporters. From these examples, we can see that the study of information diffusion plays a rather important role in advertisement and advocation.

In the literature, there are numerous works on the information diffusion over social networks, and here we summarize the works that are most representative and relevant to our study. The existing works can be classified into two categories: diffusion dynamics analysis and diffusion stability analysis. The first category focuses on analyzing the dynamic diffusion process over different kinds of networks using different mathematical models \cite{2}\nocite{3,4,5,6}-\cite{7}. Whats in \cite{2} is one of the most earlier works that studied the dynamics of information propagation through blogspace from both macroscopic and microscopic points of views. Subsequently, in \cite{3}, the authors studied how a social network affects the spread of behavior and investigated the effects of network structure on users' behavior diffusion. Rather than focusing on the behavior diffusion, Bakshy \emph{et al.} studied the role of social networks in general information diffusion through an experimental approach \cite{4}. Recently, as online social networks, e.g., Facebook and Twitter, become more and more popular, some empirical analyse were conducted using large-scale datasets, including predicting the speed and range of information diffusion on Twitter \cite{5}, modeling the global influence of a node on the rate of diffusion on Twitter \cite{6} and illustrating the statistical mechanics of rumor spreading on Facebook \cite{7}. The second category of information diffusion analysis focuses on the stability and consequence of information diffusion \cite{9}\nocite{10,11,12}-\cite{13}. In \cite{9}, the authors discussed how to extract the most influential nodes on a large-scale social network. Later, a community-based greedy algorithm was proposed for mining top-$k$ influential nodes in mobile social networks \cite{11}. How to restrain the private or contaminated information diffusion was studied in \cite{10} and \cite{12} through identifying the important information links and hubs, respectively. On the other hand, how to maximize information diffusion through a network was discussed in \cite{13} by designing effective neighbours selection strategies. In this paper, our analysis falls into the first category, i.e., to model the dynamic diffusion process.

Most of existing works on information diffusion analysis are based on machine learning methods through empirical data mining, which is based on the assumption that the training set is statistically consistent with the testing set. One conspicuous shortcoming is that the results learned from some specified dataset rely on the corresponding social network structure and may not be able to analyze or predict the future networks since the social networks are varying in a highly dynamic environment. Secondly, such machine learning based method ignores the actions and decision making of users, while the influence of users' decisions, actions and socio-economic connections on information forwarding also plays an important role in the diffusion process. Imaging that when some new information is released from one user or a group of users, whether the information can quickly diffuse or suddenly disappear largely relies on whether other users forward the information. For one user to make a decision on whether to forward or not is based on many factors, such as if the information is exciting or if his/her friends are interested on it, etc. This kind of interaction can often be modeled by using game theory \cite{Yan}. Therefore, in this paper, we propose a game theoretic framework to analyze the dynamics of information diffusion over social networks. Moreover, compared with the machine learning based method, the game theoretic one focuses on the users' behavior analysis from a microeconomics point of view, the results of which do not rely on the assumptions of a steady network structure and can be generally used to analyze and predict a future network.

In the dynamic information diffusion process, on one hand, whether one user forwards the information or not can be influenced by his/her neighbors; one the other hand, the user can also influence his/her neighbors' forwarding actions. In essence, such a dynamic mutual influence process is quite similar to the evolution process in natural ecological systems \cite{evol}, where the user with new information can be regarded the mutant and the information diffusion can be considered as the mutant spreading. Therefore, we consider the evolutionary game to model and study the dynamic information diffusion process. The proposed model reveals the dynamics of information diffusion among users through analyzing their learning, interactions and decision making. Based on the evolutionary game theoretic formulation, we analyze the dynamics of diffusion process over complete networks, uniform degree and non-uniform degree networks, with highlight of two special networks: the Erd\H{o}s-R\'enyi random network and the Barab\'asi-Albert scale-free network. We find that the dynamics of information diffusion over these three networks are scale-free and the same with each other when the network scale is sufficiently large. The experiments are conducted to validate our theoretical analysis using synthetic networks, real-world Facebook network and the Twitter hashtags dataset.

The rest of this paper is organized as follows. We first model the dynamics of information diffusion over complete networks using evolutionary game theory in Section II. Then, we model the dynamics of information diffusion over uniform degree and non-uniform degree networks using graphical evolutionary game theory in Section III and IV, respectively. Experiments results are shown in Section V and conclusions are drawn in Section VI.

\section{Diffusion Dynamics over Complete Networks}

In this section, we model the information diffusion dynamics over complete networks using evolutionary game. We first introduce the basic concepts of evolutionary game, and then elaborate the formulation of the dynamic information diffusion process.

\subsection{Basic Concepts of Evolutionary Game Theory}

The classical game theory considers a fixed set of players and focuses on the analysis of the static Nash equilibrium. The evolutionary game theory (EGT), originated from ecological biology \cite{evol}, extends the game formulation through incorporating the concept of ``population'' and emphasizes more on the dynamics of the whole population's strategies update \cite{evol2}. It has emerged as an effective modeling tool in signal processing area including peer-to-peer streaming \cite{p2p}, wireless multicast \cite{vicky}, image interpolation \cite{yanev} and adaptive filtering networks \cite{jiang1}-\cite{jiang}, as well as communications and networking area including cognitive radio networks \cite{cs}-\cite{joint} and wireless communications \cite{wireless}-\cite{tac}.

One of the most important concepts in EGT is the ``replicator dynamics'', which illustrates the dynamic process of the whole population's strategies and can provide the system state information at a particular time instance. The replicator represents the player from the population, who can reproduce his/her strategy under some specific rules of selection and mutation. EGT defines that the replicator with higher payoff can reproduce with higher rate, which is consistent with the nature selection theory. In such a case, the dynamics of the replicators can be illustrated as a set of differential equations. Consider an evolutionary game with a population of $N$ players and $M$ pure strategies $\mathcal X=\{1,2,...,M\}$. Let $n_i$ denote the number of players adopting strategy $i$ and $x_i=\frac{n_i}{N}$ denote the proportion of players adopting strategy $i$ among the whole population. In such a case, the population state can be illustrated by a vector $\mathbf x=[x_1,x_2,...,x_M]$. In EGT, the utility of a player is referred as ``fitness'' \cite{fitness}, which is defined as follows:
\begin{equation}
\Psi=(1-\alpha)\cdot B+\alpha\cdot U,\label{fitness}
\end{equation}
where $B$ is the baseline fitness representing the player's inherent property. For example, in a social network, a user's baseline fitness can be interpreted as his/her own interests on the released news. $U$ is the player's payoff which is determined by the predefined payoff matrix and the player's interactions with others. The parameter $\alpha$ represents the selection intensity, i.e., the relative contribution of the game to fitness. The case $\alpha\rightarrow 0$ represents the limit of weak selection \cite{weak}, while $\alpha\rightarrow1 $ denotes strong selection, where fitness equals payoff. Note that the selection intensity can be time variant, e.g., $\alpha(t)=\beta e^{-\epsilon t}$ which means that the contribution of game interaction decreases along with time.

According to EGT, the replicator dynamics can be defined by a set of discrete differential equations as follows:
\begin{equation}
\left\{\begin{array}{c}
\dot x_1(t)=x_1(t)\left[\overline \Psi_1(t)-\overline \Psi(t) \right],\\
\vdots\\
\dot x_i(t)=x_i(t)\left[\overline \Psi_i(t)-\overline \Psi(t) \right],\\
\vdots\\
\dot x_M(t)=x_M(t)\left[\overline \Psi_M(t)-\overline \Psi(t) \right],
\end{array}\right.\label{rd}
\end{equation}
where $\dot x_i(t)$ represents the variation of $x_i$ at time $t$, i.e., $x_i(t+1)=x_i(t)+\dot x_i(t)$, $\overline \Psi_i(t)$ represents the average fitness of players adopting strategy $i$ at time $t$ and $\overline \Psi(t)$ represents the average fitness of the whole population at time $t$. We can see that if adopting strategy $i$ can lead to a higher fitness than the average level, the proportion $x_i$ will increase and the increasing rate $\dot x_i/x_i$ is proportional to the difference between $\overline \Psi_i$ and $\overline \Psi$. Note that when the total population is sufficiently large and homogeneous, the proportion of players adopting strategy $i$ is equivalent to the probability of one individual player adopting strategy $i$, i.e., $\mathbf x$ can be interpreted as each player's mixed strategy and the replicator dynamics can be interpreted as each player's mixed strategy update.

\subsection{Evolutionary Game Formulation}

In a social network, when a series of new information are originated from one user or a small group of users, the dynamic information diffusion process heavily depends on other users' actions: to forward the information or not. For each user, whether to forward the information is determined by several factors, including the user's own interest on this information, as well as his/her neighbor's actions in the sense that if all his/her neighbors forward the information, the user may also forward the information with relatively high probability. Such a dynamic process of users' information forwarding is quite similar to the players' strategies update in the aforementioned EGT. Therefore, we can model the information diffusion dynamics over complete network using the evolutionary game as follows.

\begin{itemize}

\item \emph{Players and Population}: The users in the social networks can be regarded as players. All the users in a particular social network, i.e., Facebook or Twitter, can be regarded as a population. In addition, a large group of users in a social network can also be regarded as a population, i.e., a circle in Google plus.

\item \emph{Strategy}: In the information diffusion process, each user has two possible actions, i.e., to forward the received information or not forward, which are corresponding with two strategies as follows:
\begin{eqnarray}
\left\{\begin{array}{ll}
\bm S_f,&\mbox{forward the information}, \vspace{2mm}\\
\bm S_n,&\mbox{not forward the information}.
\end{array}\right.\label{strategy}
\end{eqnarray}

\item \emph{Payoff}: In social networks, users' payoff is determined by multiple factors, including the cost of forwarding the information, the reward obtained by forwarding/not forwarding the information (e.g, the popularity of a user in a social network or the hit rate of a website). In this paper, we model the users' payoff matrix as follows:
\begin{eqnarray}
    \begin{tabular}{ccccc}
    && $\bm S_f$ &$\bm S_n$&\vspace{1mm}\\
    \!\!\!\!$\bm S_f$ \!\!\!\!\!\!\!\!& \multirow{2}{0.01cm}{\bigg(}& $u_{ff}$&$u_{fn}$&\!\!\!\!\!\!\!\!\!\!\multirow{2}{0.01mm}{\bigg)\!\!\!\!}\\
    \!\!\!\!$\bm S_n$ \!\!\!\!\!\!\!\!& & $u_{fn}$&$u_{nn}$&
    \end{tabular}\label{payoff}
\end{eqnarray}
where a symmetric payoff structure is considered, i.e., when a user with strategy $\bm S_f$ meets a user with strategy $\bm S_n$, each of them receives the same payoff $u_{fn}$. Moreover, we assume that the payoff has been normalized within interval $(0,1)$, i.e., $0 < u_{ff},u_{fn},u_{nn} <1$. Note that under different application scenarios, the values of the payoff matrix may be different. For example, if the information is related to recent hot topics and forwarding of the information can attract more attentions from other users or website, the payoff matrix should have the following characteristic: $u_{ff}\ge u_{fn}\ge u_{nn}$. On the other hand, if the information is about useless advertisements, the payoff matrix would exhibit $u_{nn}\ge u_{fn}\ge u_{ff}$.
\end{itemize}

From the formulation, we can see that the dynamics of information diffusion over social network shares fundamental similarities with the strategy updating in the evolutionary game. In the following subsection, we will model the dynamic information diffusion process using replicator dynamics.

\subsection{Information Diffusion Dynamics over Complete Network}

In a complete network, each user has possible interactions with all other users. Under such a circumstance, once some information are released by a user, all other users are supposed to receive the information. However, whether to forward the information depends on the strategies of different users. In this scenario, we consider the network as a group of users that continuously release new information. For instance, in practical social networks, such a group of users can be a circle in the Google pluse or a group in the Facebook. Since each user in the group also connects to other users outside the group, the more users in the group forward the information, the wider the information can diffuse. Therefore, through analyzing the dynamic of users' strategies on information forwarding, we can infer how the information propagate to other users outside the group. Let us define the proportion of users adopting strategy $\bm S_f$, i.e., forward the information, as $x_f$; and the proportion of users adopting strategy $\bm S_n$ as $x_n=1-x_f$. In such a case, the network state can be described by $\mathbf x= [x_f,x_n]$.

To analyze the dynamic changing of $\mathbf x$ along with time, we discretize the dynamic information diffusion process into time slot. In each time slot, the users in the complete network are assumed to be able to observe the strategies and fitness of other users in the population. Based on the observed information, in the next time slot, each user's decision on whether forwarding the information or not is determined by which strategy can give him/her higher fitness. Thus, along with the users' strategies update slot by slot, the network state $\mathbf x$ also keeps changing slot by slot. Let us define the changing rate of the network state as the \emph{population dynamics} of information diffusion, $[\dot x_f,\dot x_n]$. According to the replicator dynamics in (\ref{rd}), the population dynamics can be modeled as follows:
\begin{align}
\dot x_f=&\ x_f\left(\overline \Psi_f-\overline \Psi\right),\label{xf}\\
\dot x_n=&\ x_n\left(\overline \Psi_n-\overline \Psi\right),\label{xn}
\end{align}
where $\overline \Psi_f$ is the average fitness of adopting strategy $\bm S_f$, i.e., forward the information, $\overline \Psi_n$ is the average fitness of adopting strategy $\bm S_n$, i.e., not forward the information, $\overline \Psi$ is the average fitness of the whole population. Since $x_f+x_n=1$, the network state can be described only by one replicator dynamics equation, i.e., either (\ref{xf}) or (\ref{xn}). From the replicator dynamics of $\dot x_f$ and $\dot x_n$, we can see that the number of users who forward the information increases if the average fitness of forwarding is above the average fitness. On the contrary, the number of users who do not forward the information increases if the average fitness of not forwarding exceeds the average fitness.

Suppose the current network state is $[x_f,1-x_f]$. For a user adopting strategy $\bm S_f$, he/she meets a user also adopting strategy $\bm S_f$ with probability $x_f$ and meets a user adopting strategy $\bm S_n$ with probability $1-x_f$. In such a case, according to the payoff matrix defined in (\ref{payoff}), the average fitness of adopting strategy $\bm S_f$, i.e., forward the information, can be calculated by
\begin{equation}
\overline \Psi_f=1-\alpha+\alpha\big[x_fu_{ff}+(1-x_f)u_{fn}\big].\label{uf}
\end{equation}
Note that we normalize the baseline fitness as 1 throughout this paper. Similarly, the average fitness of adopting strategy $\bm S_n$, i.e., not forward the information can be calculated by
\begin{equation}
\overline \Psi_n= 1-\alpha+\alpha\big[x_fu_{fn}+(1-x_f)u_{nn}\big].
\end{equation}
Then, the average fitness of the whole population can be calculated by
\begin{align}
\overline \Psi=&\ x_f\overline \Psi_f+(1-x_f)\overline \Psi_n\nonumber\\
=&\ 1\!-\!\alpha\!+\!\alpha\big[x_f^2u_{ff}+x_f(1\!-\!x_f)u_{fn}+(1\!-\!x_f)^2u_{nn}\big].\!\!\label{ua}
\end{align}
By substituting (\ref{uf})-(\ref{ua}) into (\ref{xf}), we have
\begin{align}
\dot x_f=&\ x_f(1-x_f)\left(\overline \Psi_f-\overline \Psi_n\right)\nonumber\\
=&\ \alpha x_f(1\!-\!x_f)\big[(u_{ff}\!-\!2u_{fn}\!+\!u_{nn})x_f\!+\!u_{fn}\!-\!u_{nn}\big],
\end{align}
where we can see that the selection intensity parameter $\alpha$ controls the speed of users' observation and strategy adjustment.
Thus, we can summarize the following theorem.

\emph{\textbf{Theorem 1:}} The population dynamics of information diffusion over complete networks can be described as follows:
\begin{align}
&\ \dot x_f(t)=\alpha x_f(t)(1-x_f(t))\left(a_1x_f(t)+b_1\right),\label{order} \\
&\ x_f(t+1)=x_f(t)+\dot x_f(t),
\end{align}
\begin{equation}
\!\!\!\!\!\!\!\mbox{where}\ \left\{
\begin{array}{l}
a_1= u_{ff}-2u_{fn}+u_{nn},\\
b_1= u_{fn}-u_{nn}.
\end{array}\right.
\end{equation}

\textbf{Remarks:} From the derivation of \emph{Theorem 1}, we can see that no network scale information, e.g., how many users in the network, were utilized. The information diffusion dynamics in (\ref{order}) only rely on the initial state $x_f(0)$ and the values of payoff matrix, which also shows the scale-free property.

\section{Diffusion Dynamics over Uniform Degree Networks}

In this section, we model the information diffusion dynamics over uniform degree networks, where users do not fully connect with each other. We first introduce the basic concepts of graphical evolutionary game, and then elaborate how to formulate the dynamic information diffusion process over uniform degree networks.

\subsection{Basic Concepts of Graphical Evolutionary Game Theory}

The traditional evolutionary game theory considers a population with full connections, i.e., the population is based on a complete graph. However, in many scenarios, players' spatial locations may lead to an incomplete graph structure. Graphical evolutionary game theory is introduced to study the strategies evolution in such a structured population \cite{gegreview}. In graphical EGT, in addition to the entities of players, strategy and payoff matrix, each game model is associated with a graph structure, where the vertexes represent players and the edges determine which player to interact with. Since the players only has limited connections with others, each player's fitness is locally determined from interactions with all adjacent players. In essence, the traditional evolutionary game can be regarded as a special case of graphical EGT, where the corresponding graph structure is complete. Previously, we have used graphical EGT to model the adaptive networks \cite{gegt1}, as well as the stable state of information diffusion over social networks \cite{gegt2}. The major difference is that, we focus on the dynamics analysis of information diffusion in this paper using replicator dynamics, while \cite{gegt2} focused on the final stable state of information diffusion by analyzing the evolutionarily stable state (ESS), which is also an important concept in the EGT.

\begin{figure}[!t]
\begin{minipage}[t]{0.32\linewidth}
  \centering
  \centerline{\epsfig{figure=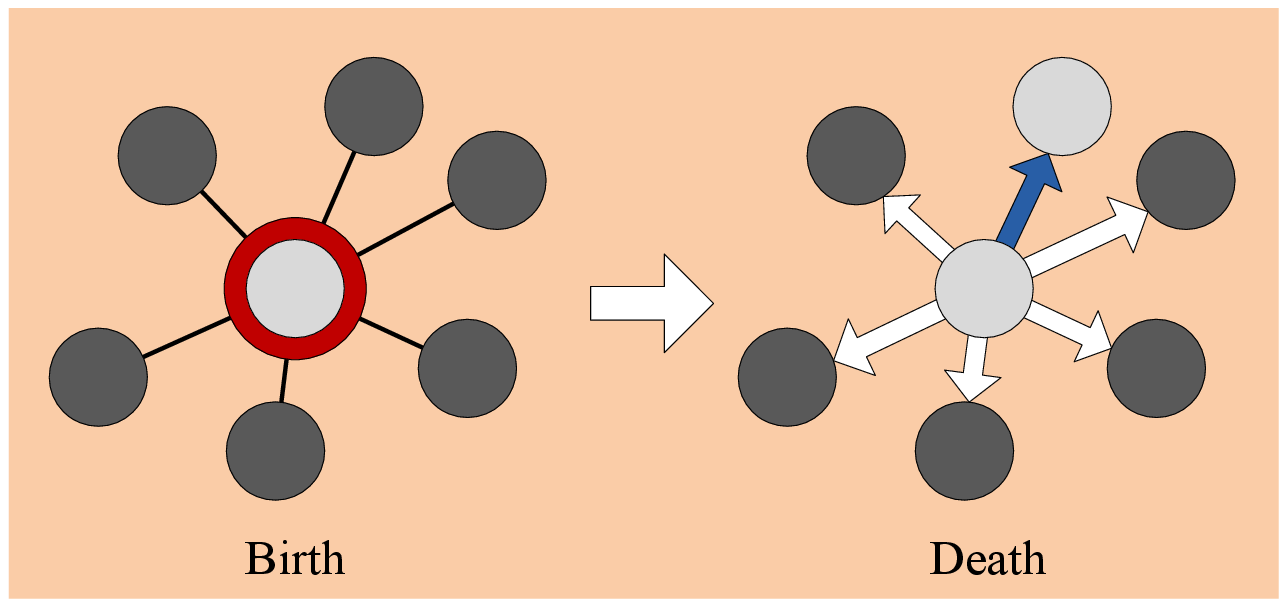,width=2.8cm}}
  \centerline{\scriptsize{(a) BD update rule.}}
\end{minipage}
\begin{minipage}[t]{0.32\linewidth}
  \centering
  \centerline{\epsfig{figure=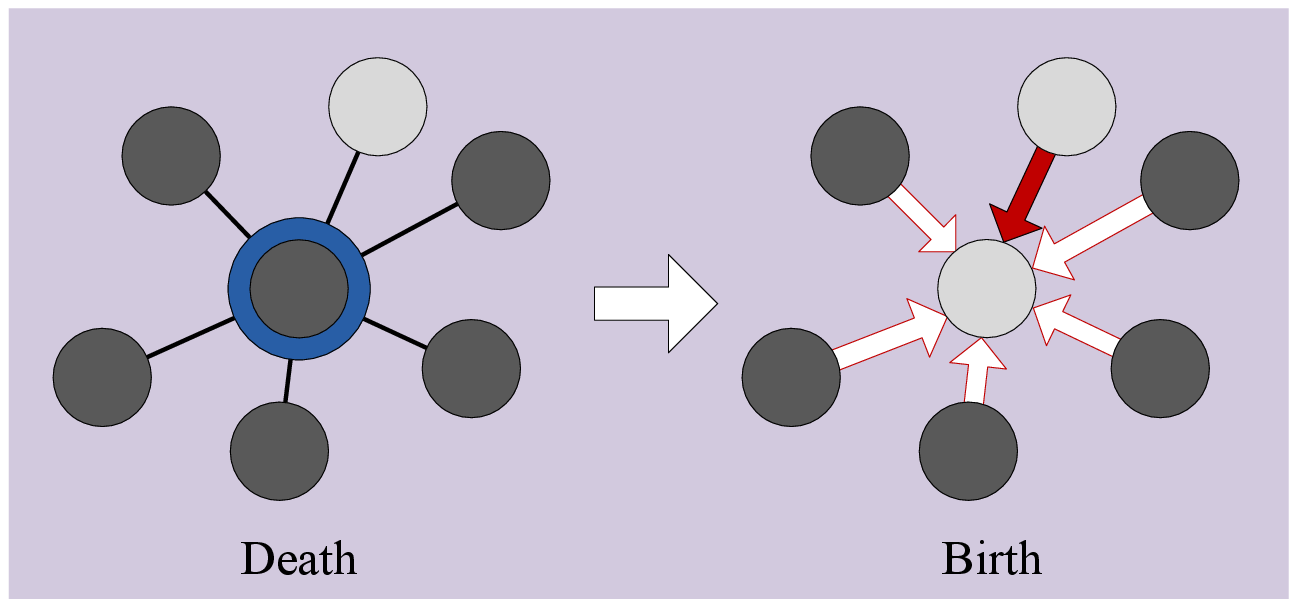,width=2.8cm}}
  \centerline{\scriptsize{(b) DB update rule.}}
\end{minipage}
\begin{minipage}[t]{0.32\linewidth}
  \centering
  \centerline{\epsfig{figure=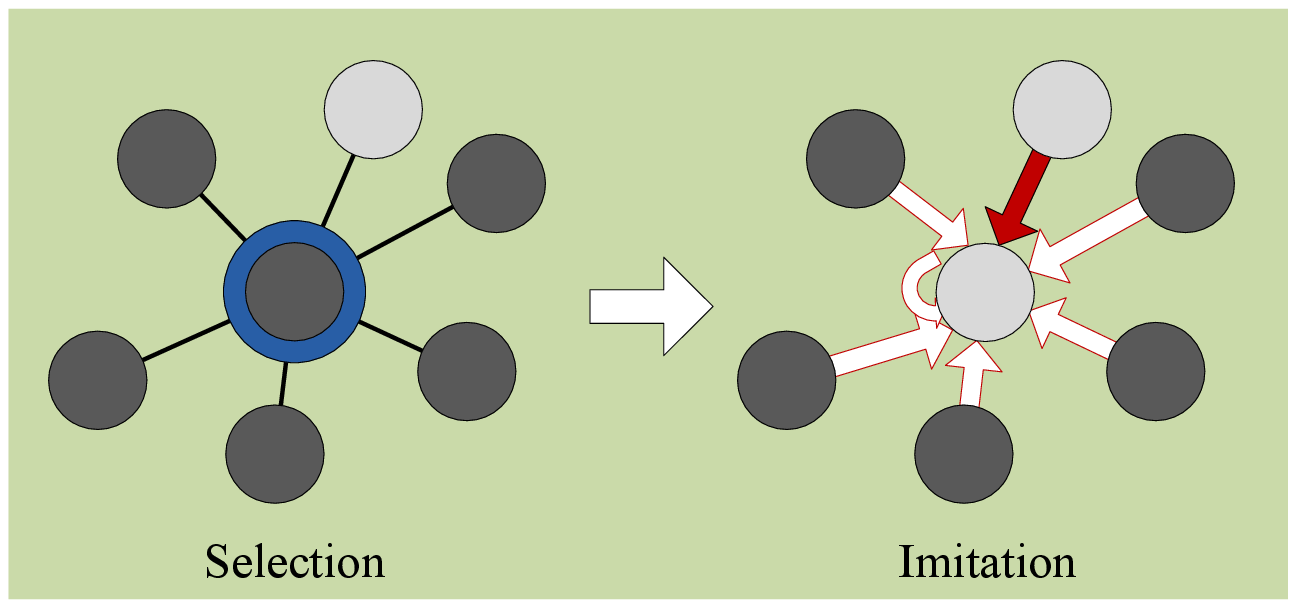,width=2.8cm}}
  \centerline{\scriptsize{(c) IM update rule.}}
\end{minipage}
\caption{Three different strategy update rules, where death selections are shown in dark blue and birth selections are shown in red.}\label{fig3}
\vspace{-5mm}
\end{figure}
Similar to that in the traditional EGT, the concept of replicator dynamics is also of importance in the graphical EGT. The difference is that it is usually analyzed under some predefined strategy updating rules, including birth-death (BD), death-birth (DB) and imitation (IM) \cite{rd}. These strategy updating rules are from the evolutionary biology field and used to model the resident/mutant evolution process as follows: (a) For BD strategy update rule, a player is chosen for reproduction with the probability being proportional to fitness (Birth process). Then, the chosen player's strategy replaces one neighbor's strategy with uniform probability (Death process), as shown in Fig.\,\ref{fig3}-(a). (b) For DB strategy update rule, a random player is chosen to abandon his/her current strategy (Death process). Then, the chosen player adopts one of his/her neighbors' strategies with the probability being proportional to their fitness (Birth process), as shown in Fig.\,\ref{fig3}-(b). (c) For IM strategy update rule, each player either adopts the strategy of one neighbor or remains with his/her current strategy, with the probability being proportional to fitness, as shown in Fig.\,\ref{fig3}-(c). In the following, we will first analyze the dynamics of information diffusion based on BD strategy update rule, and then extend the analysis to the DB and IM strategy update rules, where we will also show that these three rules are equivalent when the network degree is sufficiently large.

\subsection{Graphical Evolutionary Game Formulation}

A social network can be represented by a graph, where each node represents a user and the edge represents the relationship between users. The users can be either human in a social network or websites on the internet, while the relationship can be either friendship between users or hyperlink between webpages. When a series of new information are released by a user, the information may either diffuse over the graph or distinguish suddenly, which is determined by the user's neighbors' information forwarding actions, as well as the neighbors' neighbors' forwarding actions. Similar to the previous section, we can model the information diffusion dynamics over the graph based networks using graphical evolutionary game as follows.

\begin{itemize}

\item \emph{Graph Structure}: The social network topology is just corresponding to the graph structure of the game.

\item \emph{Players and Population, Strategy and Payoff}: The formulations are similar to those in Section II-B.
\end{itemize}

\subsection{Diffusion Dynamics over Uniform Degree Networks}

Based on the graphical evolutionary game formulation above, we analyze the information diffusion dynamics over uniform degree networks in this subsection. In the uniform scenario, an $N$-user social network based on a homogenous graph with general degree $k$ is considered. Similar to the complete network scenario, the network state of information diffusion can also be described by $\mathbf x=[x_f,1-x_f]$, where $x_f$ denotes the proportion of users who forward the information among the whole population. In this uniform degree networks scenario, our target is also to derive the dynamics of $x_f$ along with time, which reflects the diffusion scale of the information. On the other hand, unlike the complete network scenario where the probability that a player meets an a player adopting strategy $\bm S_f$ is equal to the global network state $x_f$, in a social network based on an incomplete graph, this is not necessarily true since each user only has possible connections with his/her neighbors. In such a case, due to the limited dispersal, those who adopt the same strategy, i.e., either forward the information or not, tend to form clusters. In order to take into account the correlation in strategies of two adjacent players, we define the local network states as $x_{f|f}$ and $x_{f|n}$, which represents the proportion of a user's neighbors adopting strategy $\bm S_f$, given the user is using strategy $\bm S_f$ and $\bm S_n$, respectively. In other words, $x_{f|f}$ or $x_{f|n}$ is the local network state around a user adopting strategy $\bm S_f$ or $\bm S_n$. Note that the local network state and the global network state has the relationship as follows:
\begin{align}
x_{f|f}=x_{ff}/x_f, \quad (1-x_{f|f})x_f=x_{f|n}x_n,\label{relate}
\end{align}
where $x_{ff}$ represents the global edge state, i.e., the proportion of edges on which both users adopting strategy $\bm S_f$. Similarly, we have $x_{fn}$ and $x_{nn}$, where $x_{ff}+x_{fn}+x_{nn}=1$. Thus, with the definitions of global and local network states, as well as the global edge states, we can define three dynamics of information diffusion over graph based networks as follows:
\begin{itemize}
\item \emph{Population dynamics}: $\dot x_f$, which is similar to that in the complete networks.
\item \emph{Relationship dynamics}: $\dot x_{ff}$ and $\dot x_{nn}$, which are the dynamics of global edge states and illustrate the dynamics of relationship among users. Note that $\dot x_{fn}=-\dot x_{ff}-\dot x_{nn}$.
\item \emph{Influence dynamics}: $\dot x_{f|f}$ and $\dot x_{f|n}$, which are the dynamics of local network state and illustrate the influence of one user on his/her neighbors. For instance, $\dot x_{f|f}=1$ means that all the user's neighbors adopt the same forwarding strategy with him/her, i.e., the user's neighbors are inclined to be influenced by him/her or the user is more influential. On the other hand, $\dot x_{f|n}=1$ represents an opposite characteristic.
\end{itemize}
In the following, we will analyze those dynamics of information diffusion over uniform degree networks, with the objective of deriving the close-form expression of population dynamics.

Similar to the complete network scenario in Section II, we also discretize the dynamic information diffusion process into time slot and analyze the local and global dynamics under the BD strategy updating rule. According to the BD strategy rule, in each time slot, a user is selected from the whole population with probability proportional to his/her fitness. Then, the selected user's strategy, i.e., either forward the information or not, replaces one of his/her neighbors' strategy randomly. In other words, one of the user's neighbor is influenced by the user and replicates the user's strategy. Since the user selected for reproduction probably adopts a strategy with higher fitness than the average, the physical meaning of such a dynamic strategy updating rule is equivalent to that of the replicator dynamics. Therefore, the dynamics of the network states updated under BD rule is also expected to be derived as a set of differential equations, as in (\ref{rd}). In the following derivation, we only consider the weak selection scenario, i.e., the selection intensity parameter $\alpha\rightarrow 0$. Under the weak selection, the payoff obtained from the interactions is considered as limited contribution to the overall fitness of each player, as we can see in (\ref{fitness}). Note that the results derived from weak selection often remain as valid approximations for larger selection strength \cite{weak}. Moreover, the weak selection assumption can help to achieve a close-form analysis of the dynamic information diffusion process and better reveal how the strategy diffuses over the network.

\subsubsection{Influence Dynamics and Relationship Dynamics}

Consider an $N$-user social network with uniform degree $k$. When some new information are released by a user, whether other users forward the information or not is assumed to follow the BD strategy update rule. Firstly, there is a possible instance that a user adopting strategy $\bm S_n$ is influenced by his/her neighbors adopting strategy $\bm S_f$ and deviates from his/her strategy $\bm S_n$. In a social network, this is corresponding to the scenario that if a user finds some hot information being forwarded among his/her friends, then the user would also decide to forward such kind of information. According to the BD strategy update rule, such an instance happens when a user adopting strategy $\bm S_f$ is selected for reproduction and replaces a neighbor adopting strategy $\bm S_n$. In such a case, this $(\bm S_f,\bm S_n)$ pair will switch to an $(\bm S_f,\bm S_f)$ pair. Meanwhile, the replaced neighbor, who replicates the strategy $\bm S_f$, also generates another $(k-1)x_{f|n}$ $(\bm S_f,\bm S_f)$ pairs on average. Thus, the $(\bm S_f,\bm S_f)$ pairs increase $1+(k-1)x_{f|n}$ in total if the instance occurs. Note that the instance occurs with probability $x_f(1-x_{f|f})$, where $x_f$ is the local selection probability and $1-x_{f|f}$ is the local replacement probability. Overall, we can summarize that in a unit update period, the $(\bm S_f,\bm S_f)$ pairs averagely increases
\begin{equation}
P_{i}=x_f(1-x_{f|f})\big[1+(k-1)x_{f|n}\big].\label{increase}
\end{equation}

Secondly, there is another possible instance that a user adopting strategy $\bm S_f$ is influenced by his/her neighbors adopting strategy $\bm S_n$ and deviates from his/her strategy $\bm S_f$. In a social network, this is corresponding to the scenario that if a user finds the information he/she forwarded has drawn little attention from his/her friends, then the user would decide not to forward such kind of information in the future. According to the BD strategy update rule, such an instance happens when a user adopting strategy $\bm S_n$ is selected for reproduction and replaces a neighbor adopting strategy $\bm S_f$, which leads to the $(\bm S_f,\bm S_f)$ pairs decreasing $(k-1)x_{f|f}$. In such a case, in a unit update period, the $(\bm S_f,\bm S_f)$ pairs averagely decreases
\begin{equation}
P_{d}=(1-x_f)x_{f|n}(k-1)x_{f|f}.\label{decrease}
\end{equation}

We assume that there are $N$ unit period in each time slot, i.e., there are $N$ sub-slots and only one update occurs in each sub-slot. Combining (\ref{increase}) and (\ref{decrease}), the expected changes of the global edge state $x_{ff}$ in one time slot, i.e., the relationship dynamics $\dot x_{ff}$, can be written as
\begin{align}
\dot x_{ff}=&\ \frac{Nx_f(1-x_{f|f})\big[1+(k-1)x_{f|n}\big]}{Nk/2}-\nonumber\\
&\ \frac{N(1-x_f)x_{f|n}(k-1)x_{f|f}}{Nk/2},\label{denom}
\end{align}
where the denominator $Nk/2$ is the total number of edges in the network.
Similarly, we can have the other relationship dynamics $\dot x_{nn}$ as follows:
\begin{align}
\dot x_{nn}=&\ \frac{N(1-x_f)\big[1+(k-1)x_{n|f}\big]}{Nk/2}-\nonumber\\
&\ \frac{Nx_f(1-x_{f|f})(k-1)x_{n|n}}{Nk/2}.\label{denom2}
\end{align}

According to \cite{rd}, Ohtsukia and Nowak found that the local network states $x_{f|f}$ and $x_{f|n}$ change with a rate of order $1$, while the global network state $x_f$ changes with a rate of order $\alpha$, which controls the speed of dynamics. In such a case, due to the weak selection, the local network states will converge to equilibria in a much faster rate than the global network state. Therefore, we have a separation of two time scales, and thus, the global network state $x_f$ can be regarded as constant during the convergence of influence dynamics $\dot x_{f|f}$ and $\dot x_{f|n}$. This is because the dynamics of local network state are only in terms of a local area, which contains at most $k$ users. Therefore, in such a small scale, the local dynamics can change and converge quite fast. However, the dynamics of global network state are associated with all users, i.e., the whole networks, the dynamics would be much slower. Based on the relationship dynamics in (\ref{denom}) and (\ref{denom2}), the influence dynamics $\dot x_{f|f}$ and $\dot x_{f|n}$ can be derived as follows:
\begin{align}
&\dot x_{f|f}=\frac{\dot x_{ff}}{x_f}\nonumber\\
&\!\!= \frac{x_f(1-x_{f|f})\big[1\!+\!(k-1)x_{f|n}\big]\!-\!(1-x_f)x_{f|n}(k-1)x_{f|f}}{kx_f/2}\nonumber\\
&\!\!= \frac{(1-x_{f|f})\big[1+(k-1)x_{f|n}\big]-(1-x_{f|f})(k-1)x_{f|f}}{k/2}\nonumber\\
&\!\!= \frac{2}{k}\Big\{1\!+\!(k-1)\big[x_{f|f}x_{f|f}\!+\!x_{f|n}\big(1\!-\!x_{f|f})\big]\!-\!kx_{f|f}\Big\},
\end{align}
and
\begin{align}
&\dot x_{f|n}=1-\frac{\dot x_{nn}}{1-x_f}\nonumber\\
&\!\!= \frac{2}{k}\Big\{(k-1)\big[x_{f|f}x_{f|n}+x_{f|n}(1-x_{f|n})\big]-kx_{f|n}\Big\}.
\end{align}
By setting $\dot x_{f|n}=\dot x_{f|f}=0$ and using the relationships described in (\ref{relate}), we can obtain the stable points (equilibrium) of the influence dynamics as
\begin{align}
x^*_{f|n}= \frac{(k-2)x_f}{k-1},\quad
x^*_{f|f}= \frac{(k-2)x_f+1}{k-1}.\label{xffs}
\end{align}

\subsubsection{Population Dynamics}\label{twocases}

Given the dynamics of local network states, i.e., the influence dynamics, we can derive the population dynamics by analyzing users' global strategy update. There are two cases that can lead to the dynamics of global network state $x_f$ as follows.
\begin{itemize}
\item \emph{Case 1:} If a user's strategy is updated from $\bm S_n$ to $\bm S_f$, the global network state $x_f$ will increase a unit.

\item \emph{Case 2:} if a user's strategy is updated from $\bm S_f$ to $\bm S_n$, the global network state $x_f$ will decrease a unit.
\end{itemize}
According to the BD strategy update rule, a user changing his/her strategy is due to the strategy replacement (replication) of his/her neighbor, who is with higher fitness and selected for reproduction. In such a case, \emph{Case 1} occurs when a user with strategy $\bm S_f$ is chosen for reproduction and one of his/her neighbors, who was adopting strategy $\bm S_n$, replicates the strategy $\bm S_f$. In other words, one of the user's neighbors is influenced by him/her through observation and interaction, and changes the strategy from $\bm S_n$ to $\bm S_f$. On the other hand, \emph{Case 2} occurs when a user with strategy $\bm S_n$ is chosen for reproduction and one of his/her neighbors, who was adopting strategy $\bm S_f$, replicates the strategy $\bm S_n$. In other words, one of the user's neighbors is influenced by him/her and changes the strategy from $\bm S_f$ to $\bm S_n$. Through analyzing the occurrence probability of \emph{Case 1} and \emph{Case 2}, we can calculate the expected changing of the global state $x_f$ in each time slot, which is just the population dynamics, as shown in the following theorem.

\emph{\textbf{Theorem 2:}} The population dynamics of information diffusion over uniform degree networks under Birth-Death strategy update rule can be described as follows:
\begin{align}
&\ \dot x_f(t)=\frac{\alpha (k-2)}{(k-1)} x_f(t)(1-x_f(t))\left(a_2x_f(t)+b_2\right),\label{order2} \\
&\ x_f(t+1)=x_f(t)+\dot x_f(t),
\end{align}
\begin{equation}
\mbox{where}\ \left\{
\begin{array}{l}
a_2= (k-2)(u_{ff}-2u_{fn}+u_{nn}),\\
b_2= u_{ff}+(k-2)u_{fn}-(k-1)u_{nn}.
\end{array}\right.
\end{equation}
\begin{proof}
See Appendix \ref{prooftheorem2}.
\end{proof}

\textbf{Remarks:} From \emph{Theorem 2}, we can see that the form of population dynamics of information diffusion over uniform degree networks is quite similar to that over complete network in (\ref{order}). The dynamics in (\ref{order2}) only rely on the initial state $x_f(0)$, the values of payoff matrix and the degree of the network, regardless of the network scale information. Therefore, the population dynamics of information diffusion over uniform degree networks also shows the scale-free property. Moreover, in real social networks, the degree of each user usually exhibits that $k\gg 2$. In such a case, (\ref{order2}) can be further approximated by
\begin{align}
\!\!\!\dot x_{f}= &\ \frac{\alpha (k-2)^2}{(k-1)}x_f(1-x_f)\big[(u_{ff}-2u_{fn}+u_{nn})x_f\nonumber\\
&\ +\frac{u_{ff}-u_{nn}}{k-2}+u_{fn}-u_{nn}\big]\nonumber\\
=&\ \alpha^\prime x_f(1\!-\!x_f)\big[(u_{ff}\!-\!2u_{fn}\!+\!u_{nn})x_f\!+\!u_{fn}\!-\!u_{nn}\big],
\end{align}
where $\alpha^\prime=\frac{\alpha (k-2)^2}{(k-1)}$. We can see that, the population dynamics of information diffusion over uniform degree networks are exactly same as that over complete network as in (\ref{order}). This is because, in a uniform degree network with sufficiently large degree, i.e., each user is with sufficiently large number of neighbors, the information forwarding strategy of one user is influenced by a large number of other users, which is similar to that in the complete networks. In essence, the complete network is a special case of the uniform degree networks when $k\rightarrow N$. Moreover, such a phenomenon also validates that the dynamics derived by the BD strategy update rule is equivalent with the replicator dynamics in complete networks. Up to now, \emph{Theorem 2} shows the dynamics of information diffusion under BD strategy update rule. The following theorem will show the relationship between the dynamics of information diffusion under BD, DB and IM strategy update rules.

\emph{\textbf{Theorem 3:}} The population dynamics of information diffusion over uniform degree networks under Birth-Death strategy update rule, Death-Birth strategy update rule and Imitation strategy update rule are equivalent when the network degree is sufficiently large.
\begin{proof}
See Appendix \ref{prooftheorem4}.
\end{proof}

\section{Diffusion Dynamics over Non-uniform Degree Networks}

In this section, we extend the analysis in last section to the non-uniform degree networks scenario. In the non-uniform scenario, we consider an $N$-user social network based on a graph whose degree exhibits distribution $\lambda(k)$. This distribution means that when randomly choosing one user on the network, the probability of the chosen user with $k$ neighbors is $\lambda(k)$. In such a case, the average degree of the network is
\begin{equation}
\overline k=\sum_{k=0}^{+\infty}\lambda(k)k.\label{ak}
\end{equation}
Note that we do not take degree correlation into account, i.e., the degrees of all users are independent of each other. Similar to the analysis in Section III, the information diffusion dynamics over non-uniform degree networks can also be modeled by graphical evolutionary game and the formulation is no different with that in Section III-B except the graph structure. In the following, we first analyze the general case of information diffusion dynamics over non-uniform degree networks under BD strategy update rule. Then, we highlight two special cases, i.e., two kinds of typical networks, Erd\H{o}s-R\'enyi random network and Barab\'asi-Albert scale-free networks.

\subsection{General Case}

The analytical method of the non-uniform scenario is same with that in Section III-C, i.e., first deriving the relationship and influence dynamics and finding the corresponding equilibrium, and then deriving the population dynamics. Therefore, we can follow the same derivations in Section III-C to derive the population dynamics of information diffusion over non-uniform networks. While the difference is that the distribution of users' degrees should be taken into account. According to the BD strategy update rule, there are two kinds of users: the selected user and the replaced neighbor. It is obvious that the degree of the selected user obeys distribution $\lambda(k)$. However, the replaced neighbor does not since if a pair is selected by random, the degree distribution of the user on the specific pair is not $\lambda(k)$ but rather $\frac{k\lambda(k)}{\sum_{k=0}^{+\infty}k\lambda(k)}$ \cite{degreec}. Through taking different expectations with respect to the different kinds of users' degree, we can derive the population dynamics of information diffusion over non-uniform degree networks in the following theorem.

\emph{\textbf{Theorem 4:}} The population dynamics of information diffusion over non-uniform degree networks under Birth-Death strategy update rule can be described as follows:
\begin{align}
\!\!\!\!&\dot x_f(t)\!=\!\frac{\alpha ({\overline k}\!-\!1)({\overline {k^2}}\!-\!2{\overline k})}{({\overline {k^2}}-{\overline k})^2} x_f(t)(1\!-\!x_f(t))\left(a_3x_f(t)\!+\!b_3\right),\label{order3}\!\!\! \\
\!\!\!&x_f(t+1)=x_f(t)+\dot x_f(t),
\end{align}
\begin{equation}
\mbox{where}\left\{
\begin{array}{l}
a_3= ({\overline {k^2}}-2{\overline k})(u_{ff}-2u_{fn}+u_{nn}),\\
b_3= {\overline k}u_{ff}+({\overline {k^2}}-2{\overline k})u_{fn}-({\overline {k^2}}-{\overline k})u_{nn}.
\end{array}\right.
\end{equation}
\begin{proof}
See Appendix \ref{prooftheorem3}.
\end{proof}

\textbf{Remarks:} From \emph{Theorem 4}, we can see that the information diffusion dynamics over non-uniform degree networks maintain the similar form as in \emph{Theorem 1} and \emph{Theorem 2}. However, the scale-free property may not hold due to the term $\overline {k^2}=\sum_{k=0}^{+\infty}k^2\lambda(k)$, which is the expectation of $k^2$ and may contain the network scale information. Moreover, in \cite{gegt2}, we analyzed the evolutionary stable states of information diffusion under IM strategy update rule. Here, in \emph{Theorem 4}, we can also find three stable states $0$, $1$ and $\frac{b_3}{a_3}$ in (\ref{order3}), which is consistent with the results in \cite{gegt2}.

\subsection{Two Special Cases}
In this subsection, we discuss two special cases of the non-uniform degree networks, Erd\H{o}s-R\'enyi random network \cite{er} and Barab\'asi-Albert scale-free network \cite{ba}. For the Erd\H{o}s-R\'enyi random (ER) network, the degree follows a Poisson distribution, i.e.,
\begin{equation}
\lambda_{\mbox{\scriptsize ER}}(k)=\frac{e^{-\overline k}\overline k^k}{k!}\ \mbox{ and }\ \overline {k^2}=\overline k(\overline k+1).
\end{equation}
In such a case, according to \emph{Theorem 4}, the population dynamics of information diffusion over Erd\H{o}s-R\'enyi random network are
\begin{align}
\dot x_f^{\mbox{\scriptsize ER}}= &\ \alpha\left(\frac{ {\overline k}-1}{\overline k}\right)^2x_f^{\mbox{\scriptsize ER}}(1-x_f^{\mbox{\scriptsize ER}})\nonumber\\
&\ \big[(\overline k-1)(u_{ff}-2u_{fn}+u_{nn})x_f^{\mbox{\scriptsize ER}}+\nonumber\\
&\ \ \ u_{ff}+(\overline k-1)u_{fn}-\overline ku_{nn}\big].\label{ernetwork}
\end{align}
When the average degree of the network $\overline k\gg 1$, $\dot x_f^{\mbox{\scriptsize ER}}$ in (\ref{ernetwork}) can be approximated by
\begin{align}
\!\!\!\!\dot x_f^{\mbox{\scriptsize ER}}=\alpha_{\mbox{\scriptsize ER}} x_f^{\mbox{\scriptsize ER}}(1\!-\!x_f^{\mbox{\scriptsize ER}})\big[(u_{ff}\!-\!2u_{fn}\!+\!u_{nn})x_f\!+\!u_{fn}\!-\!u_{nn}\big],\!\!
\end{align}
where $\alpha_{\mbox{\scriptsize ER}}=\alpha\frac{ ({\overline k}-1)^3}{\overline k^2}\approx\alpha \overline k$. We can see that the population dynamics in Erd\H{o}s-R\'enyi random network shares exactly the same form with that in complete network and uniform degree network (when the uniform degree $k\gg 2$). Moreover, since $\alpha$ is adjustable, by taking $\alpha=\frac{\alpha^\prime}{\overline k}$, we can see that $\dot x_f^{\mbox{\scriptsize ER}}$ would also become scale-free.

For the Barab\'asi-Albert scale-free (BA) network, the degree follows a power law distribution, i.e.,
\begin{equation}
\lambda_{\mbox{\scriptsize BA}}(k)\varpropto k^{-\xi}\ \mbox{ and }\ \overline {k^2}\dot =\overline k^2\log N/4\ (\mbox{when }\xi=3).
\end{equation}
In such a case, according to \emph{Theorem 4}, the information diffusion dynamics of Barab\'asi-Albert scale-free network are
\begin{align}
\dot x_f^{\mbox{\scriptsize BA}}= &\ \frac{\alpha ({\overline k}-1)({\overline k\log N/4}-2)}{({\overline k\log N/4}-1)^2} x_f^{\mbox{\scriptsize BA}}(1- x_f^{\mbox{\scriptsize BA}})\cdot\nonumber\\
&\ \big[(\overline k\log N/4-2)(u_{ff}-2u_{fn}+u_{nn}) x_f^{\mbox{\scriptsize BA}}+u_{ff}+\nonumber\\
&\ \ \ (\overline k\log N/4-2)u_{fn}-(\overline k\log N/4-1)u_{nn}\big].\label{banetwork}
\end{align}
When the network scale $N$ is sufficiently large, $\dot x_f^{\mbox{\scriptsize BA}}$ in (\ref{banetwork}) can be approximated by
\begin{align}
\!\!\!\!\dot x_f^{\mbox{\scriptsize BA}}=\alpha_{\mbox{\scriptsize BA}} x_f^{\mbox{\scriptsize BA}}(1\!-\!x_f^{\mbox{\scriptsize BA}})\big[(u_{ff}\!-\!2u_{fn}\!+\!u_{nn})x_f^{\mbox{\scriptsize BA}}\!+\!u_{fn}\!-\!u_{nn}\big],\!\!\label{scalefreeapp}
\end{align}
where $\alpha_{\mbox{\scriptsize BA}}=\frac{\alpha ({\overline k}-1)({\overline k\log N/4}-2)^2}{({\overline k\log N/4}-1)^2}\approx \alpha ({\overline k}-1)$. Thus, the population dynamics in Barab\'asi-Albert scale-free network also shares exactly the same form with that in complete network, and by taking $\alpha=\frac{\alpha^\prime}{\overline k-1}$, $\dot x_f^{\mbox{\scriptsize BA}}$ would also become scale-free.

\section{Experiments}

In this section, we conduct experiments to verify the information diffusion dynamics analysis. First, we simulate the information diffusion process on synthetic networks and real-world network, i.e., the Facebook social network to verify our theoretical analysis by setting different payoff matrices. Then, we use the Twitter hashtags dataset to estimate the payoff matrix corresponding to each hashtag by fitting the curve of real-world information diffusion process.

\subsection{Synthetic Networks and Real-World Network}

In the experiment of synthetic networks, we generate four kinds of networks to simulate the information diffusion process:
\begin{itemize}
\item the complete network;
\item the uniform-degree network;
\item the Erd\H{o}s-R\'enyi random network;
\item the Barab\'asi-Albert scale-free network.
\end{itemize}
For each network, we generate 1000 users, where only one user adopts strategy $\bm S_f$ and all other users adopt $\bm S_n$. In the simulations, four kinds of payoff matrices are considered:
\begin{itemize}
\item \emph{Case 1:} $u_{ff}>u_{fn}>u_{nn}$
\begin{eqnarray}
u_{ff}=0.8;\quad u_{fn}=0.6;\quad u_{nn}=0.4.
\end{eqnarray}
\item \emph{Case 2:} $u_{fn}>u_{ff}>u_{nn}$
\begin{eqnarray}
u_{ff}=0.6;\quad u_{fn}=0.8;\quad u_{nn}=0.4.
\end{eqnarray}
\item \emph{Case 3:} $u_{fn}>u_{nn}>u_{ff}$
\begin{eqnarray}
u_{ff}=0.4;\quad u_{fn}=0.8;\quad u_{nn}=0.6.
\end{eqnarray}
\item \emph{Case 4:} $u_{nn}>u_{fn}>u_{ff}$
\begin{eqnarray}
u_{ff}=0.4;\quad u_{fn}=0.6;\quad u_{nn}=0.8.
\end{eqnarray}
\end{itemize}

\begin{figure*}[!t]
\begin{minipage}[t]{0.24\linewidth}
  \centering
  \centerline{\epsfig{figure=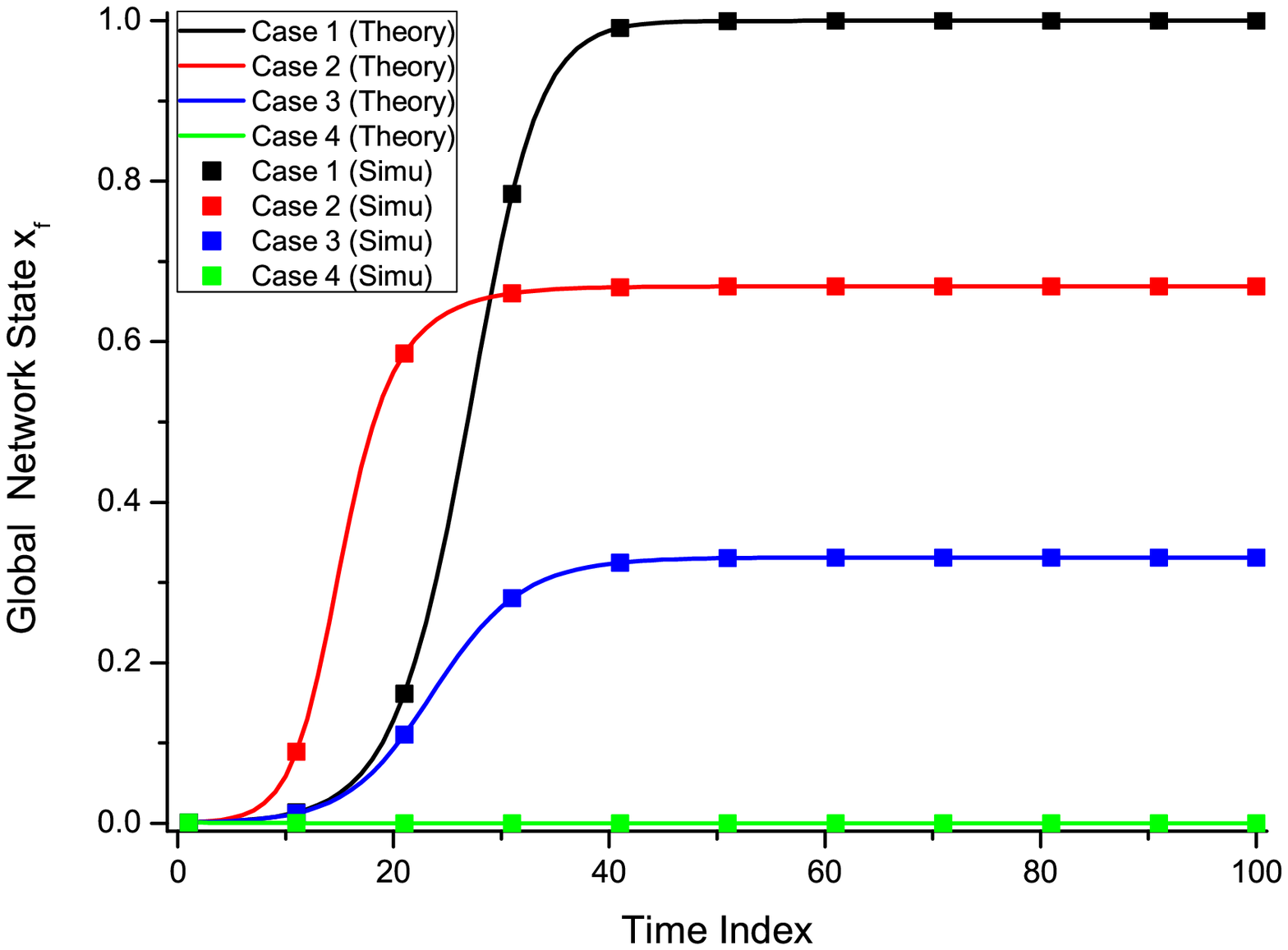,width=4.3cm}}
  \vspace{-0.1cm}
  \center{\scriptsize{(a) Complete network.}}
\end{minipage}
\hfill
\begin{minipage}[t]{0.24\linewidth}
  \centering
  \centerline{\epsfig{figure=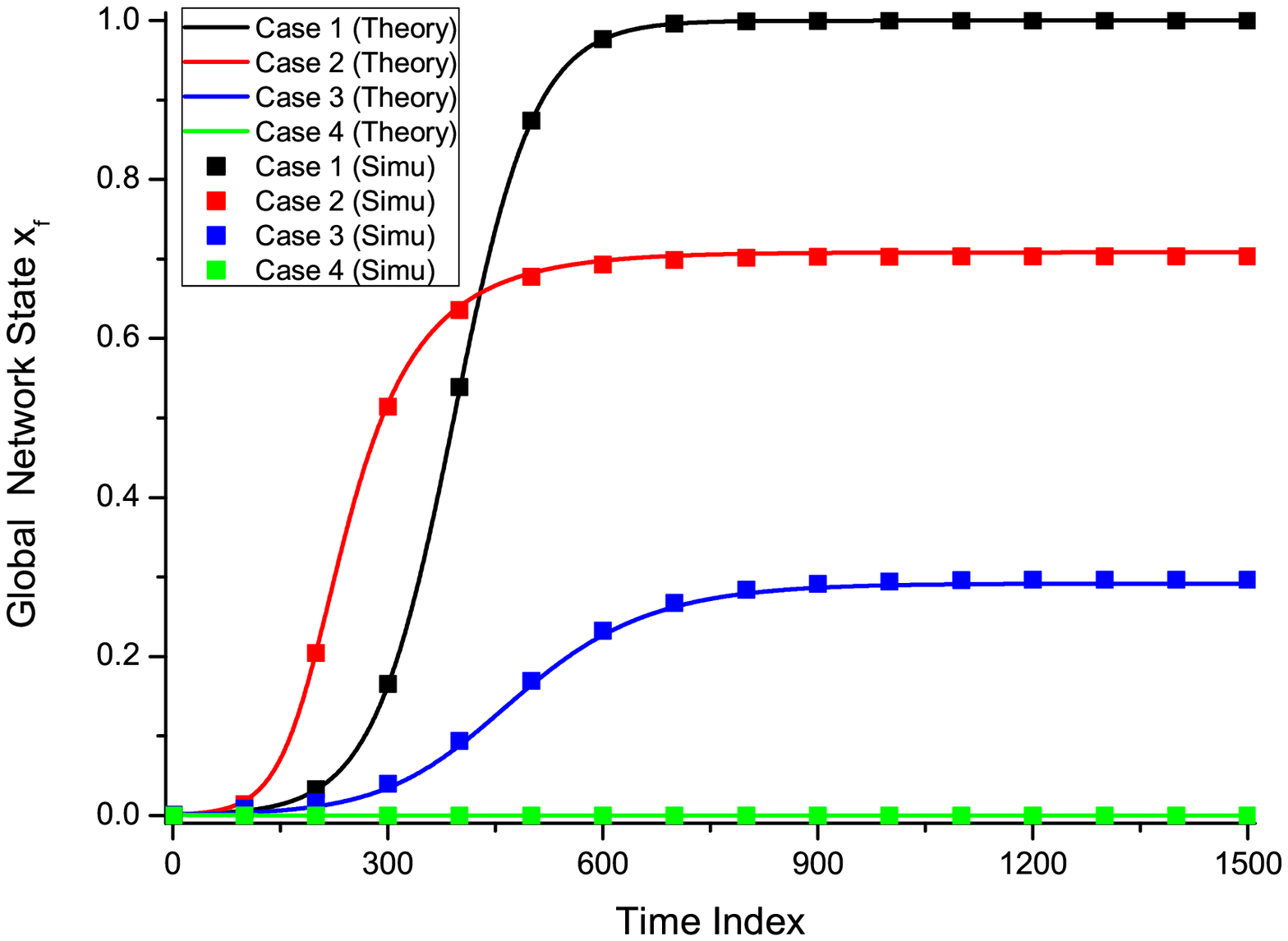,width=4.3cm}}
  \vspace{-0.1cm}
  \center{\scriptsize{(b) Uniform degree network.}}
\end{minipage}
\begin{minipage}[t]{0.24\linewidth}
  \centering
  \centerline{\epsfig{figure=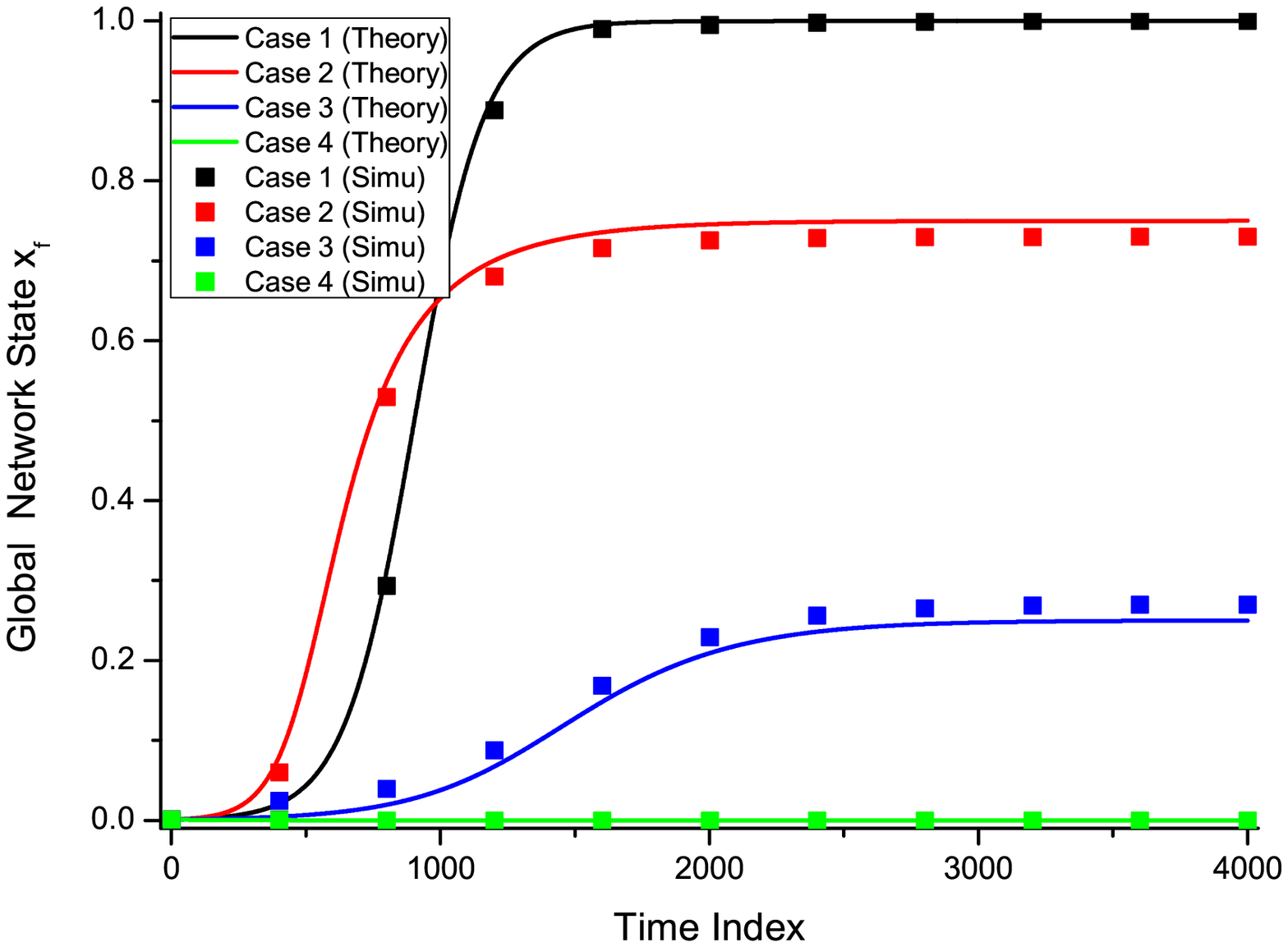,width=4.3cm}}
  \vspace{-0.1cm}
  \center{\scriptsize{(c) Erd\H{o}s-R\'enyi random network.}}
\end{minipage}
\hfill
\begin{minipage}[t]{0.24\linewidth}
  \centering
  \centerline{\epsfig{figure=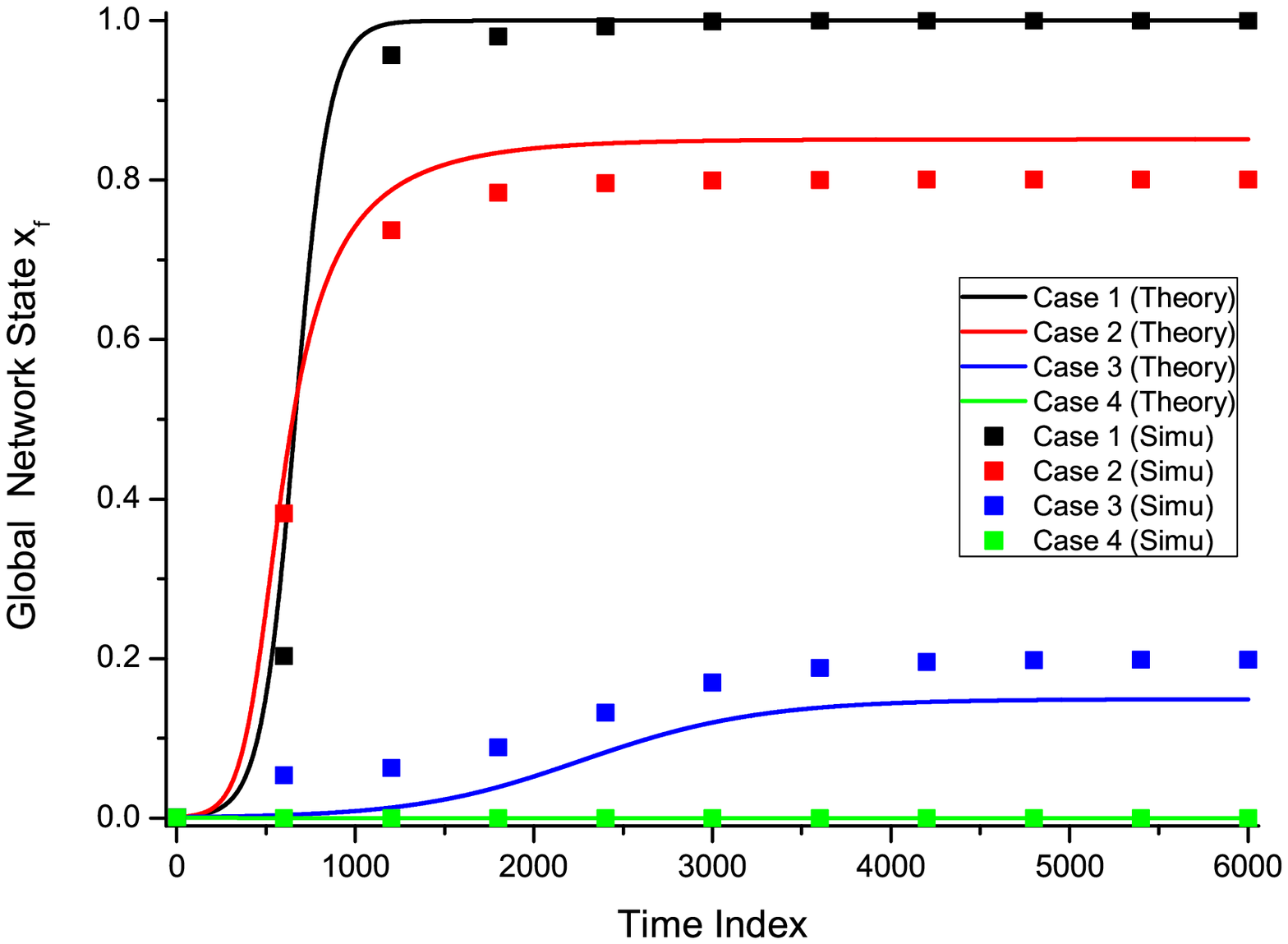,width=4.3cm}}
  \vspace{-0.1cm}
  \center{\scriptsize{(d) Barab\'asi-Albert scale-free network.}}
\end{minipage}
\caption{Simulation results for synthetic networks.}\label{simur}
\vspace{-3mm}
\end{figure*}

Fig.\,\ref{simur}-(a) and (b) shows the experiment results for the complete and uniform-degree networks under different payoff matrices. The theoretical results are calculated from \emph{Theorem 1} and \emph{Theorem 2} directly, while the simulation results are obtained by simulating the BD strategy update rule over the generated network. For each simulation run, the strategy update steps are repeated until the global network state $x_f$ converges. Meanwhile, the network structure is re-generated every $500$ runs to prevent any spurious results based on one particular realization of a specific network type. From Fig.\,\ref{simur}-(a) and (b), we can see that all the simulation results are consistent with the theoretical results, which verifies the correctness of the conclusions in \emph{Theorem 1} and \emph{Theorem 2}. Moreover, different settings of the payoff matrix can lead to different dynamics of information diffusion. For \emph{Case 1} where $u_{ff}>u_{fn}>u_{nn}$, the globe network state tends to 1 which means that all users would forward the information since forwarding can obtain higher utility than not forwarding. On the contrary, for \emph{Case 4} where $u_{nn}>u_{fn}>u_{ff}$, no one would forward the information. For \emph{Case 2} and \emph{Case 3}, the results show that a portion of users would forward the information which is determined by the relationship between $u_{ff}$ and $u_{nn}$, i.e., if $u_{ff}>u_{nn}$ more than half would forward the information and if $u_{ff}<u_{nn}$ less than half would forward the information.

Fig.\,\ref{simur}-(c) and (d) show the experiment results for the non-uniform degree networks under different payoff matrices, including the Erd\H{o}s-R\'enyi random network and the Barab\'asi-Albert scale-free network. The theoretical results are calculated from (\ref{ernetwork}) and (\ref{banetwork}) directly, while the simulation results are obtained by simulating the BD strategy update rule over the two generated networks. We can see that the all the simulation results agree well with the theoretical results. In Fig.\,\ref{simur}-(d), the gap for the Barab\'asi-Albert scale-free networks is due to the fact that there is weak dependence between the global network state and the network degree, while we neglected such dependence in the diffusion analysis. Similarly, different settings of payoff matrices can lead to different information diffusion dynamics, which are determined by the relationships between $u_{ff}$, $u_{fn}$ and $u_{nn}$.

\begin{figure}[!t]
  \centerline{\epsfig{figure=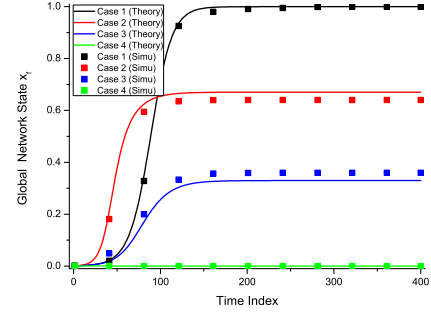,width=5.5cm}}
  \caption{Simulation results for the real-world Facebook network.}\label{facebook}\vspace{-5mm}
\end{figure}

In the experiment of real-world network, we evaluate the information diffusion process over Facebook social network \cite{facebookgrpah}. The Facebook dataset contains totally 4039 users and 88234 edges, and  the average degree is about 40 \cite{dataset}. Fig.\,\ref{facebook} shows the experiment results under different payoff matrix settings. The theoretical results are calculated by \emph{Theorem 2}, while the simulation results are obtained by simulating the BD strategy update rule over the Facebook graph. It can be seen that the simulation results match well with the theoretical results, while the small gaps are mainly due to the neglected dependence between the global network state and the network degree. Therefore, the experiment results verify the correctness of our theoretical analysis on the real-world network.

\subsection{Twitter Hashtags Dataset Evaluation}

In the previous two subsections, we first setup the users' payoff matrix, and then conducted experiments to verify the dynamics of information diffusion over different kinds of networks. In this subsection, a reverse process is conducted, which means that we use the Twitter hashtag dataset to estimate the payoff matrices corresponding to different hashtags by fitting the curves as shown in Fig.\,\ref{twitter}. Note that the curves in Fig.\,\ref{twitter} are corresponding to the $\dot x_f(t)$ in our model since the vertical axis is the increased mention times of one hashtag in an hour. The Twitter hashtag dataset contains the the number of mention times per hour of 1000 Twitter hashtags with corresponding time series, which are the 1000 hashtags with highest total mention times among 6 million hashtags from Jun. to Dec. 2009 \cite{dataset}.

\begin{figure*}[!t]
\begin{minipage}[t]{0.24\linewidth}
  \centering
  \centerline{\epsfig{figure=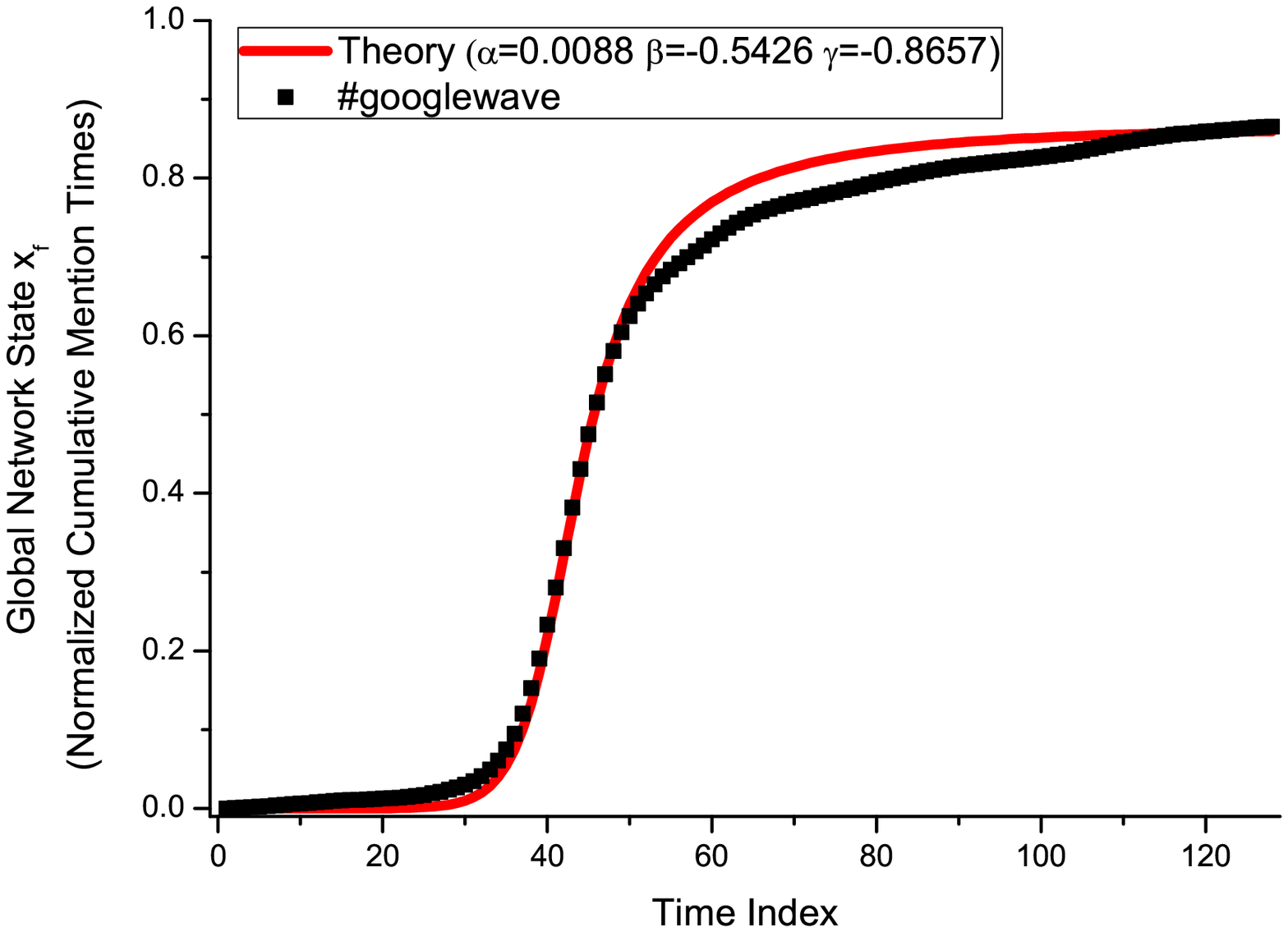,width=4.3cm}}
  \vspace{-0.1cm}
  \center{\scriptsize{(a) googlewave.}}
\end{minipage}
\hfill
\begin{minipage}[t]{0.24\linewidth}
  \centering
  \centerline{\epsfig{figure=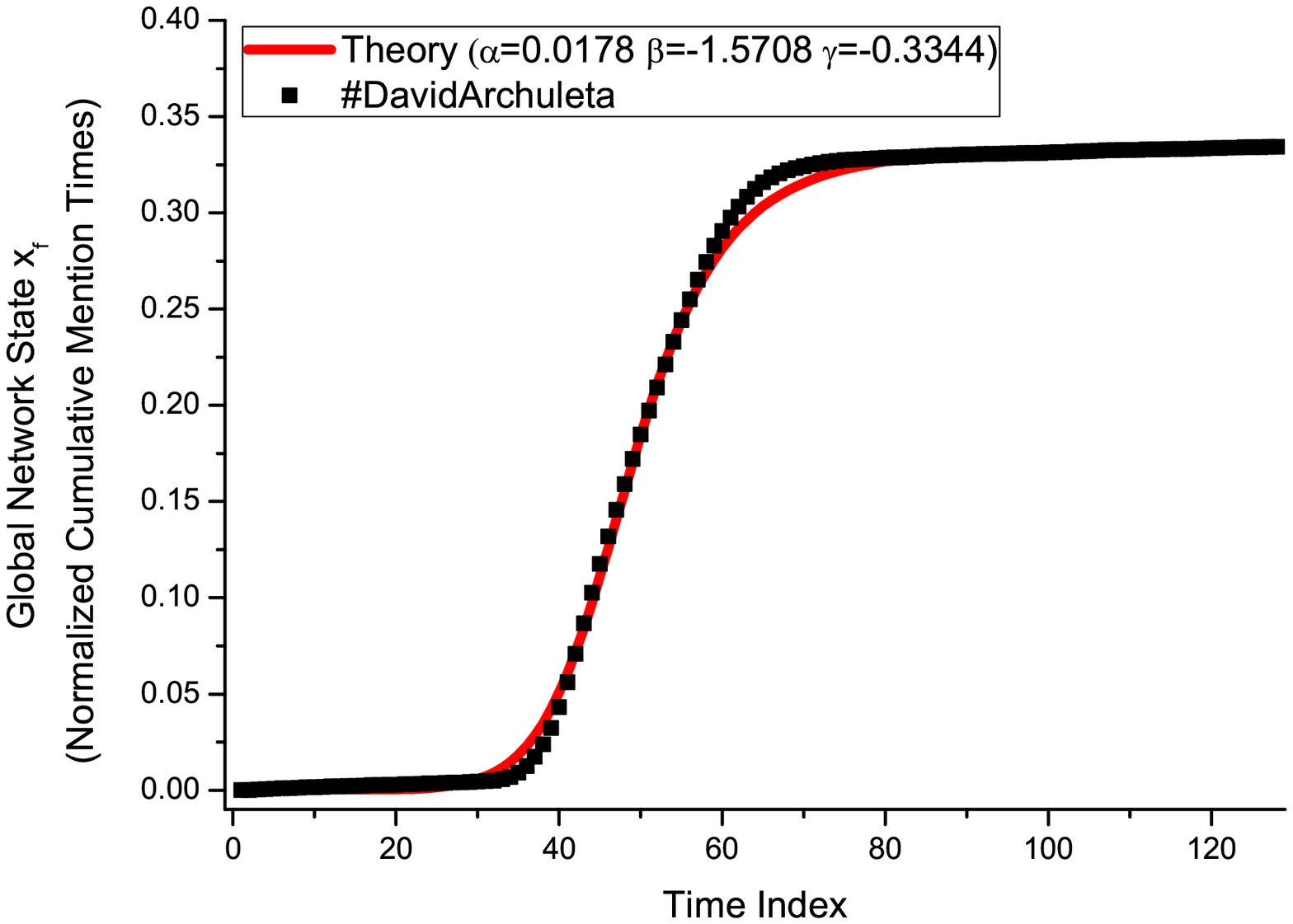,width=4.3cm}}
  \vspace{-0.1cm}
  \center{\scriptsize{(b) DavidArchuleta.}}
\end{minipage}
\begin{minipage}[t]{0.24\linewidth}
  \centering
  \centerline{\epsfig{figure=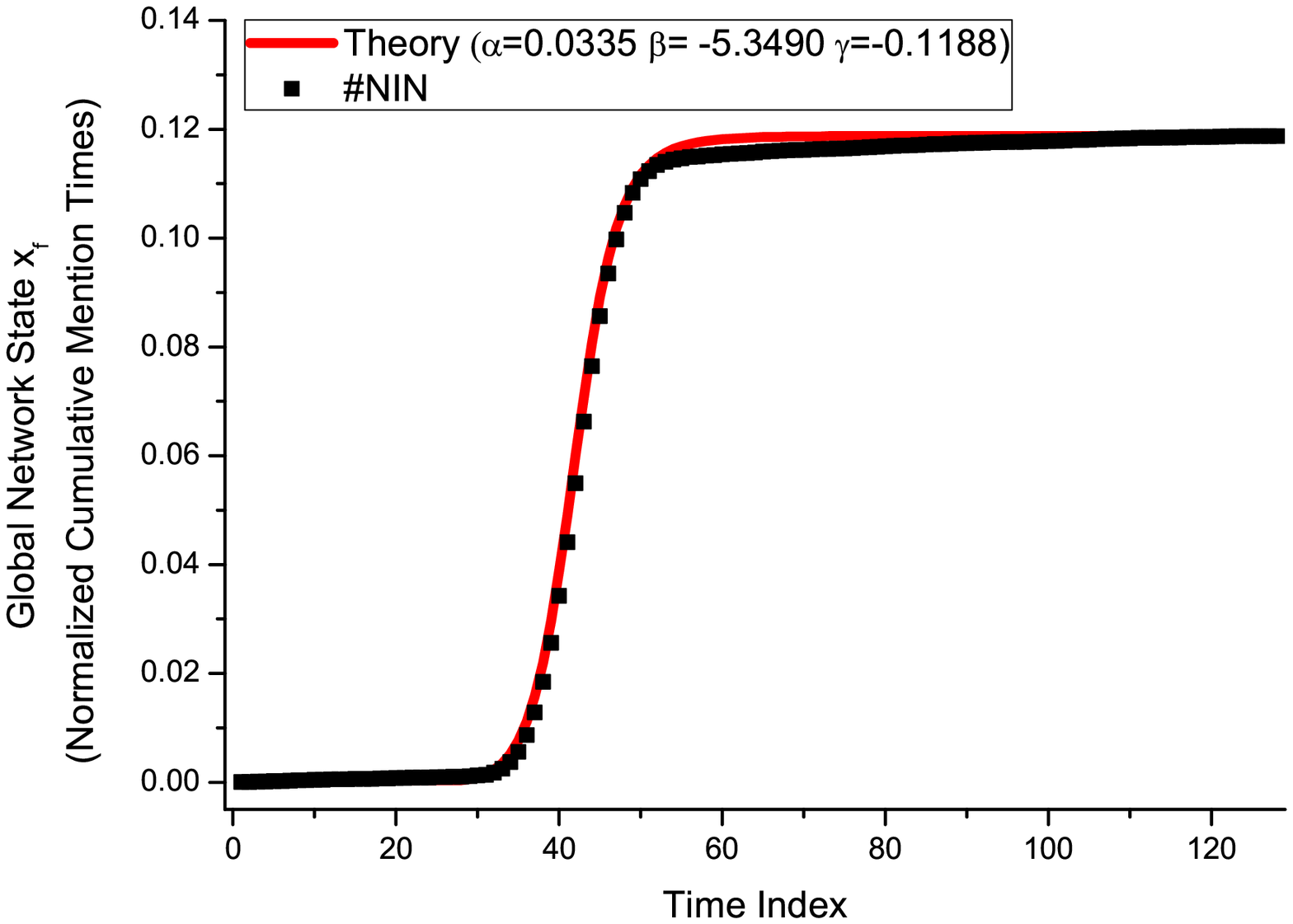,width=4.3cm}}
  \vspace{-0.1cm}
  \center{\scriptsize{(c) NIN.}}
\end{minipage}
\hfill
\begin{minipage}[t]{0.24\linewidth}
  \centering
  \centerline{\epsfig{figure=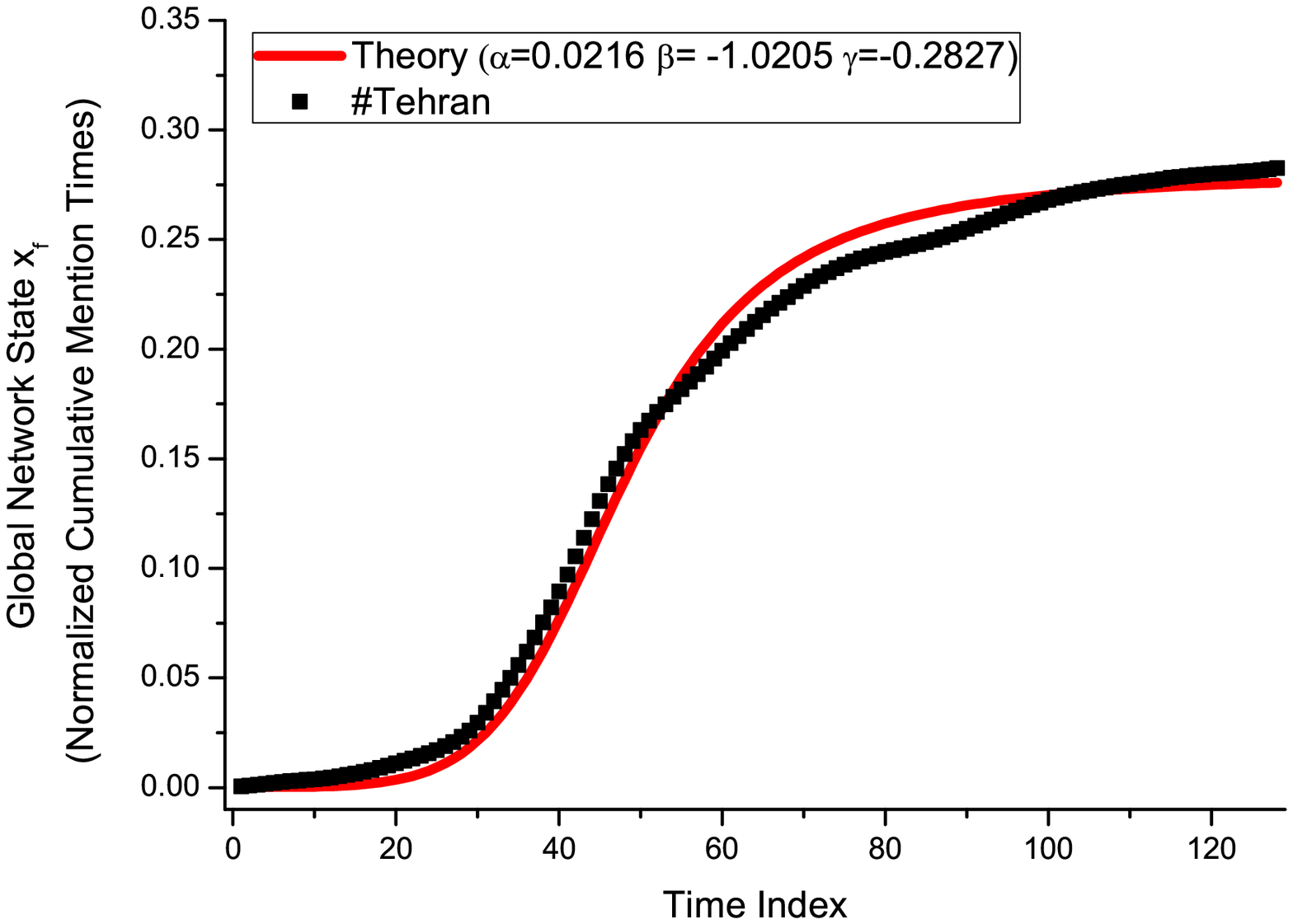,width=4.3cm}}
  \vspace{-0.1cm}
  \center{\scriptsize{(d) Tehran.}}
\end{minipage}
\caption{The curve fitting of different hashtags diffusion dynamics.}\label{xfsimu}\vspace{-5mm}
\end{figure*}

Let us first derive the closed-form expression for the global network state $x_f(t)$. In \emph{Theorem 1}, \emph{Theorem 2} and \emph{Theorem 4}, we can see that the dynamics of information diffusion over three kinds of networks share the same form as follows:
\begin{align}
\frac{d x_f}{d t}=\beta e^{-\epsilon t} x_f(1-x_f)\left( x_f+\gamma\right) ,\label{general}
\end{align}
where $\alpha=e^{-\epsilon t}$ is considered as time-variance and different kinds of networks have different coefficients $\beta$ and $\gamma$. Using the separation of variables method, we can derive the implicit closed-form expression of $x_f$ as follows:
\begin{align}
\frac{(\gamma+1)\mbox{ln}x_f-\gamma\mbox{ln}(1-x_f)+\mbox{ln}(-x_f-\gamma)}{\gamma(\gamma+1)}=-\frac{\beta}{\epsilon}e^{-\epsilon t}+c,\label{close}
\end{align}
where $c$ is a constant and can be calculated by the initial condition $x_f(t=0)$. In such a case, we can estimate the parameters $\epsilon$, $\beta$ and $\gamma$ using (\ref{close}) through fitting the Twitter hashtag dataset. Fig.\,\ref{xfsimu} shows the curve fitting results of four hashtags using least squares method, where the vertical axis is global network state $x_f(t)$. The mention times of different hashtags per hour in the Twitter dataset are first normalized within interval $[0,1]$ and then accumulated over time to get the cumulative mention times as shown by solid black square. From the figure, we can see that our model can fit the real-world information diffusion data very well, which means that the global network state of information diffusion can be accurately predicted by the proposed evolutionary game theoretic model.

\begin{figure}[!t]
  \centerline{\epsfig{figure=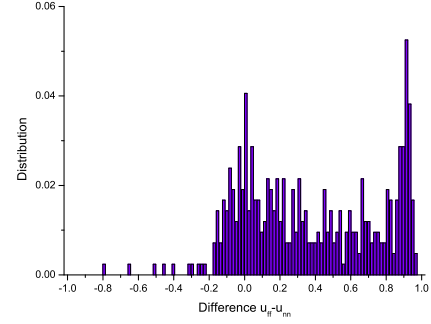,width=5.5cm}}
  \caption{Distribution of $u_{ff}-u_{nn}$ among all 1000 hashtags.}\label{distr}\vspace{-5mm}
\end{figure}

Considering that the Twitter social network also exhibits the scale-free phenomenon and the network scale is sufficiently large, we can obtain the relationship of the payoff matrix according to (\ref{scalefreeapp}) by setting $\alpha_{\mbox{\scriptsize BA}}=\frac{e^{-\epsilon t}}{\overline k-1}$ as follows:
\begin{align}
u_{ff}-2u_{fn}+u_{nn}=&\ \beta,\\
u_{fn}-u_{nn}=&\ \beta\gamma.
\end{align}
If $u_{fn}$ is normalized as 1, then we can calculate $u_{ff}$ and $u_{nn}$ through solving the equation set above. The difference between $u_{ff}$ and $u_{nn}$ can reflect the popularity of a hashtag, i.e., $u_{ff}-u_{nn}>0$ means high popularity since forwarding the information can obtain higher utility than not forwarding; on the other hand, $u_{ff}-u_{nn}<0$ means low popularity. Fig.\,\ref{distr} shows the distribution of $u_{ff}-u_{nn}$ among all $1000$ hashtags in the Twitter dataset. We can see that most of hashtags are with high popularity, which is because those 1000 hashtags are with highest total mention times among 6 million hashtags from Jun. to Dec. 2009. Therefore, using such an analytical method, the information over social network can be categorized into different levels according to the popularity of them. On the other hand, it can also be used to category users in the social network. Consider that if the sport-related information always show high popularity in a group of users, those users probably share the same interests on the sports, which means that delivering them with sport-related advertisements would be more effective.

\begin{figure*}[!t]
\begin{minipage}[t]{0.24\linewidth}
  \centering
  \centerline{\epsfig{figure=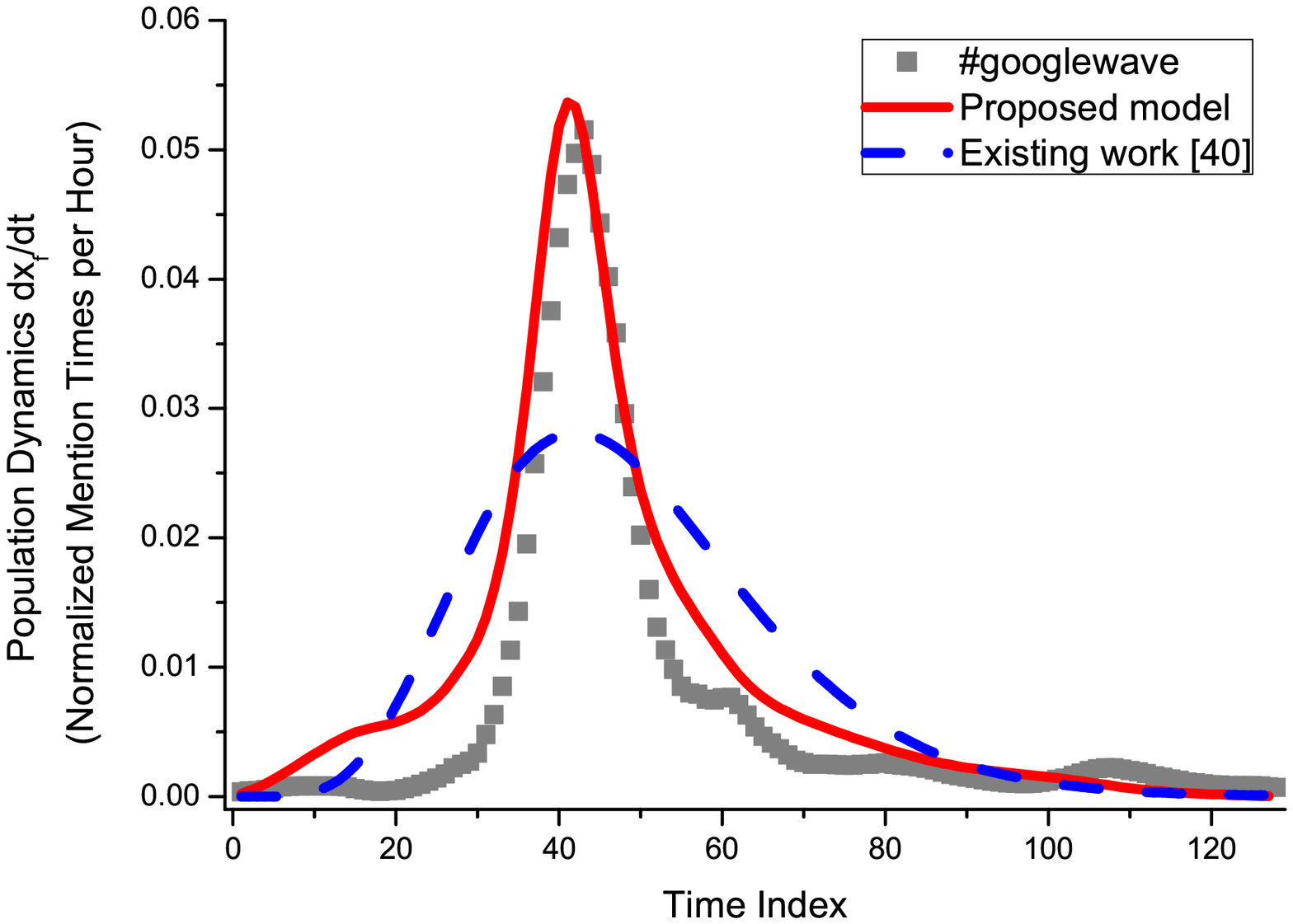,width=4.3cm}}
  \vspace{-0.1cm}
  \center{\scriptsize{(a) googlewave.}}
\end{minipage}
\hfill
\begin{minipage}[t]{0.24\linewidth}
  \centering
  \centerline{\epsfig{figure=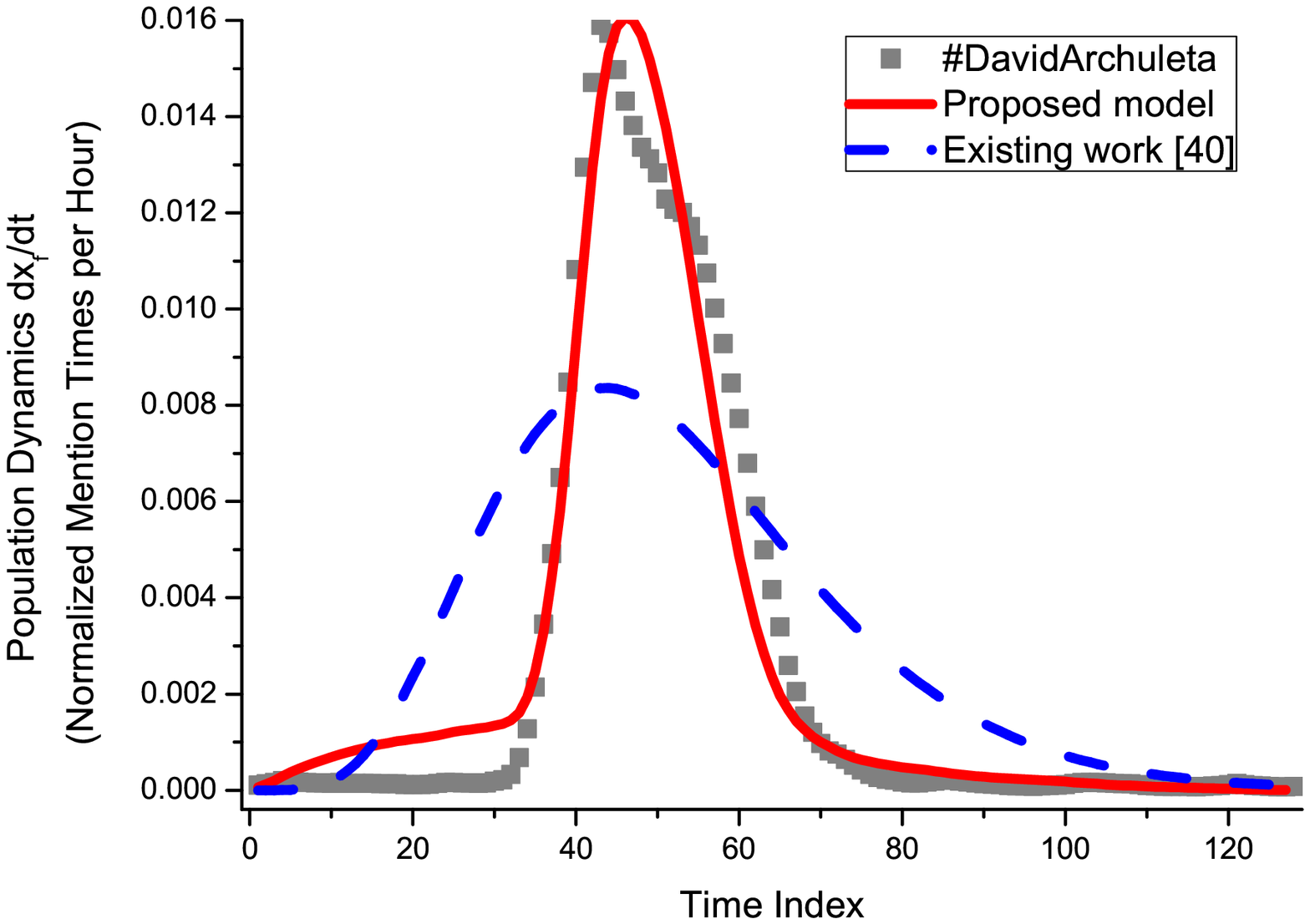,width=4.3cm}}
  \vspace{-0.1cm}
  \center{\scriptsize{(b) DavidArchuleta.}}
\end{minipage}
\begin{minipage}[t]{0.24\linewidth}
  \centering
  \centerline{\epsfig{figure=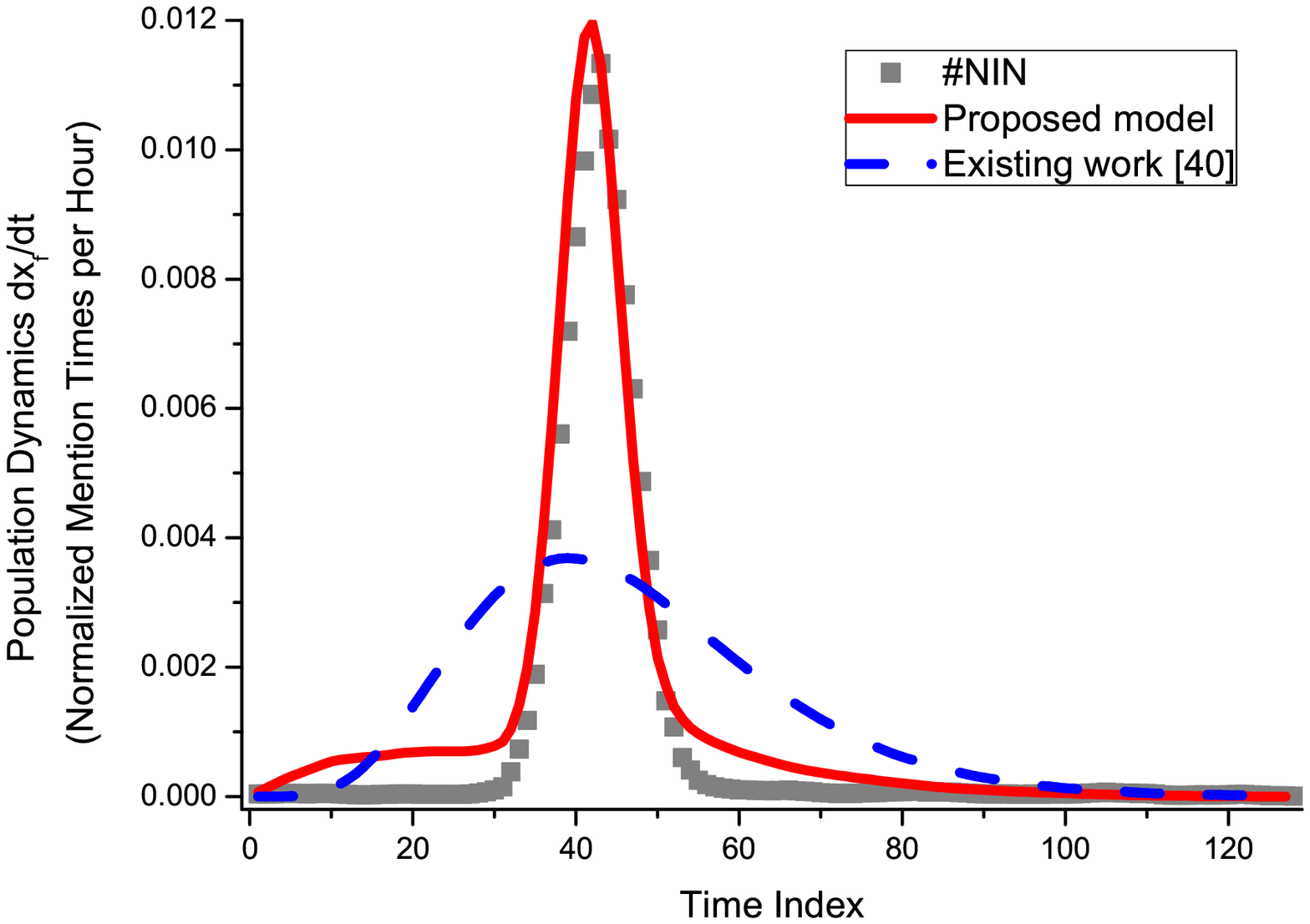,width=4.3cm}}
  \vspace{-0.1cm}
  \center{\scriptsize{(c) NIN.}}
\end{minipage}
\hfill
\begin{minipage}[t]{0.24\linewidth}
  \centering
  \centerline{\epsfig{figure=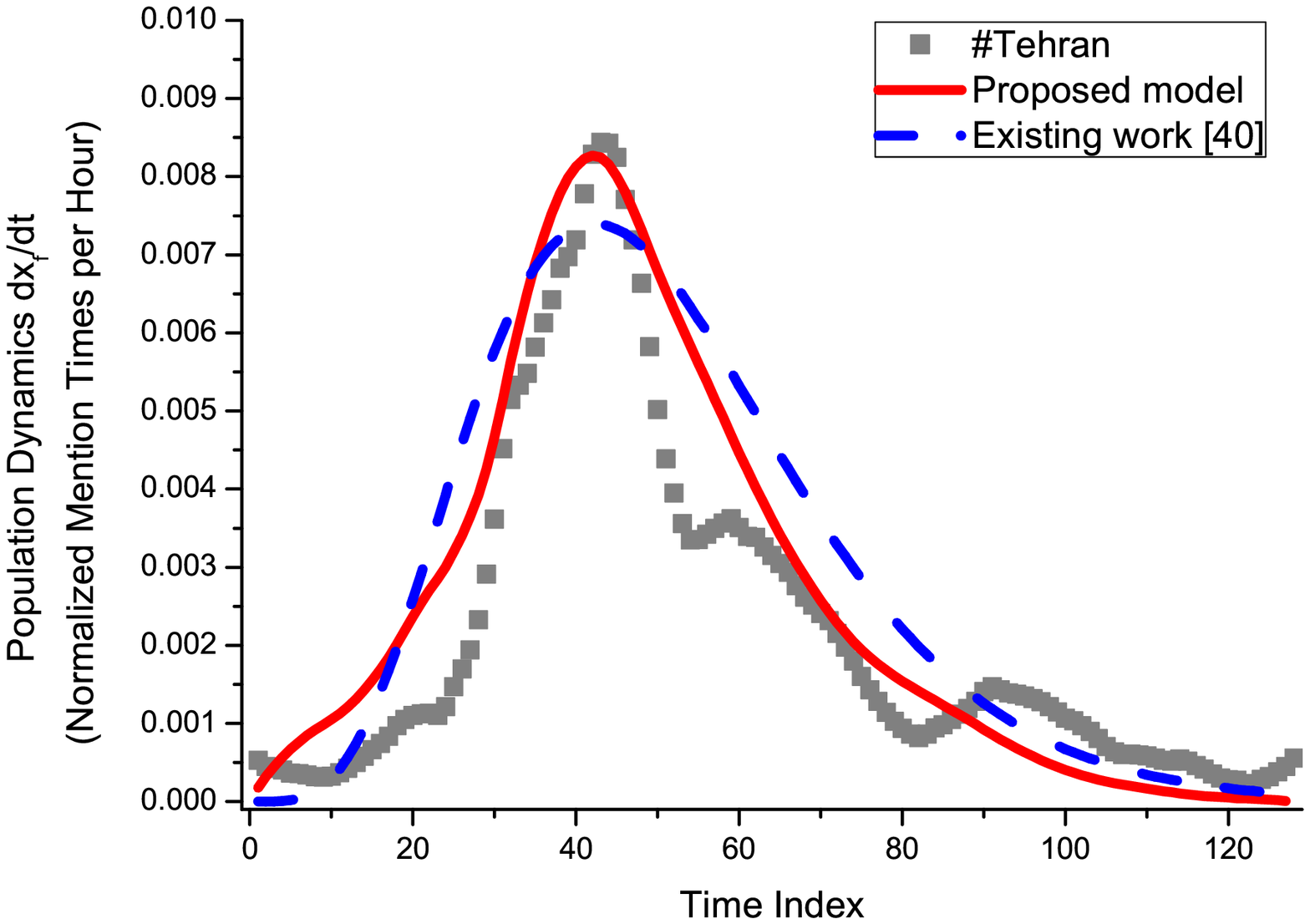,width=4.3cm}}
  \vspace{-0.1cm}
  \center{\scriptsize{(d) Tehran.}}
\end{minipage}
\caption{Comparison with the existing work.}\label{comp}\vspace{-5mm}
\end{figure*}

Based on the estimated payoff matrix, we can further simulate the dynamics $\dot x_f(t)$ using our proposed model. In this experiment, we compare our results with one of the most related exsiting works \cite{meme2} using data mining method, in which the dynamics of information diffusion are predicted by the following model
\begin{align}
\frac{d x_f}{d t}=q_1t^{q_2}e^{-q_3 t},
\end{align}
where the parameters $q_1$, $q_2$ and $q_3$ can also be estimated through least-squares curve fitting in a similar way. Fig.\,\ref{comp} shows the comparison results, where the vertical axis is the dynamics $\dot x_f(t)$ and the mention times of different hastags per hour in the Twitter dataset are normalized within interval $[0,1]$ and denoted by solid black square. From the figure, we can see that our model can fit the real-world information diffusion dynamics better than the data mining method in \cite{meme2} since the users' interactions and decision making behaviours are taken into account.

\begin{figure*}[!t]
\begin{minipage}[t]{0.24\linewidth}
  \centering
  \centerline{\epsfig{figure=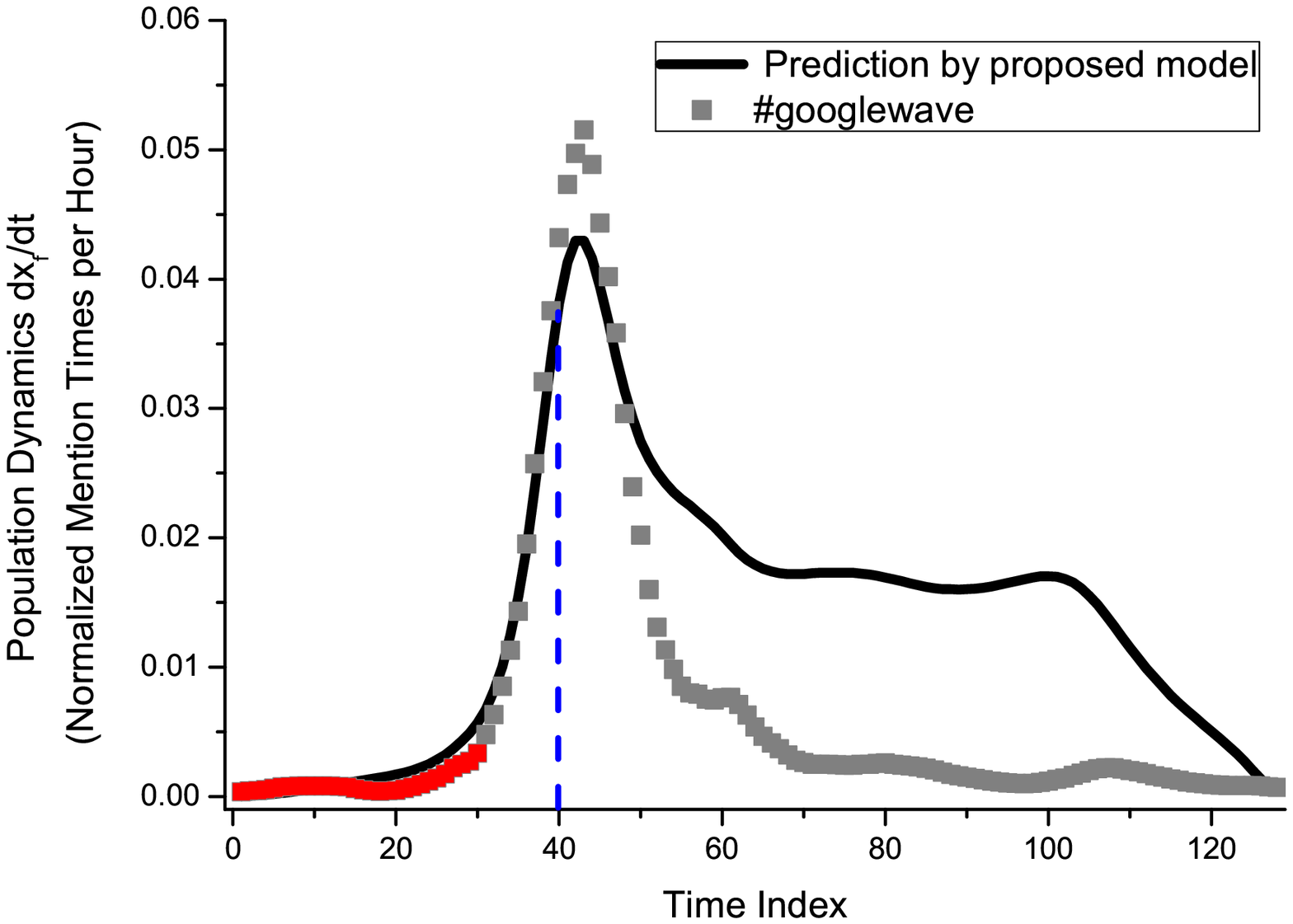,width=4.3cm}}
  \vspace{-0.1cm}
  \center{\scriptsize{(a) Curve fitting using 25\% data.}}
\end{minipage}
\hfill
\begin{minipage}[t]{0.24\linewidth}
  \centering
  \centerline{\epsfig{figure=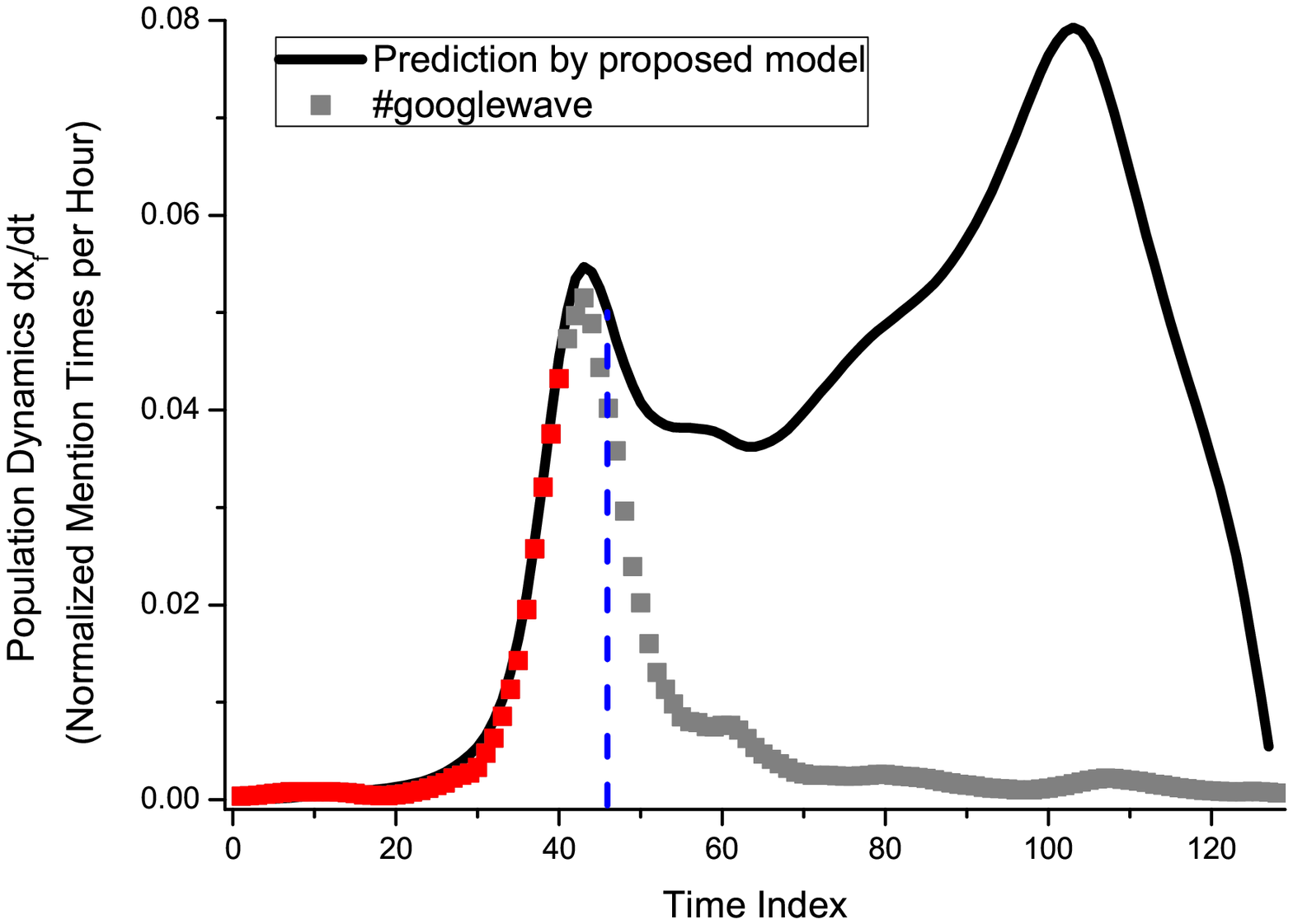,width=4.3cm}}
  \vspace{-0.1cm}
  \center{\scriptsize{(b) Curve fitting using 30\% data.}}
\end{minipage}
\begin{minipage}[t]{0.24\linewidth}
  \centering
  \centerline{\epsfig{figure=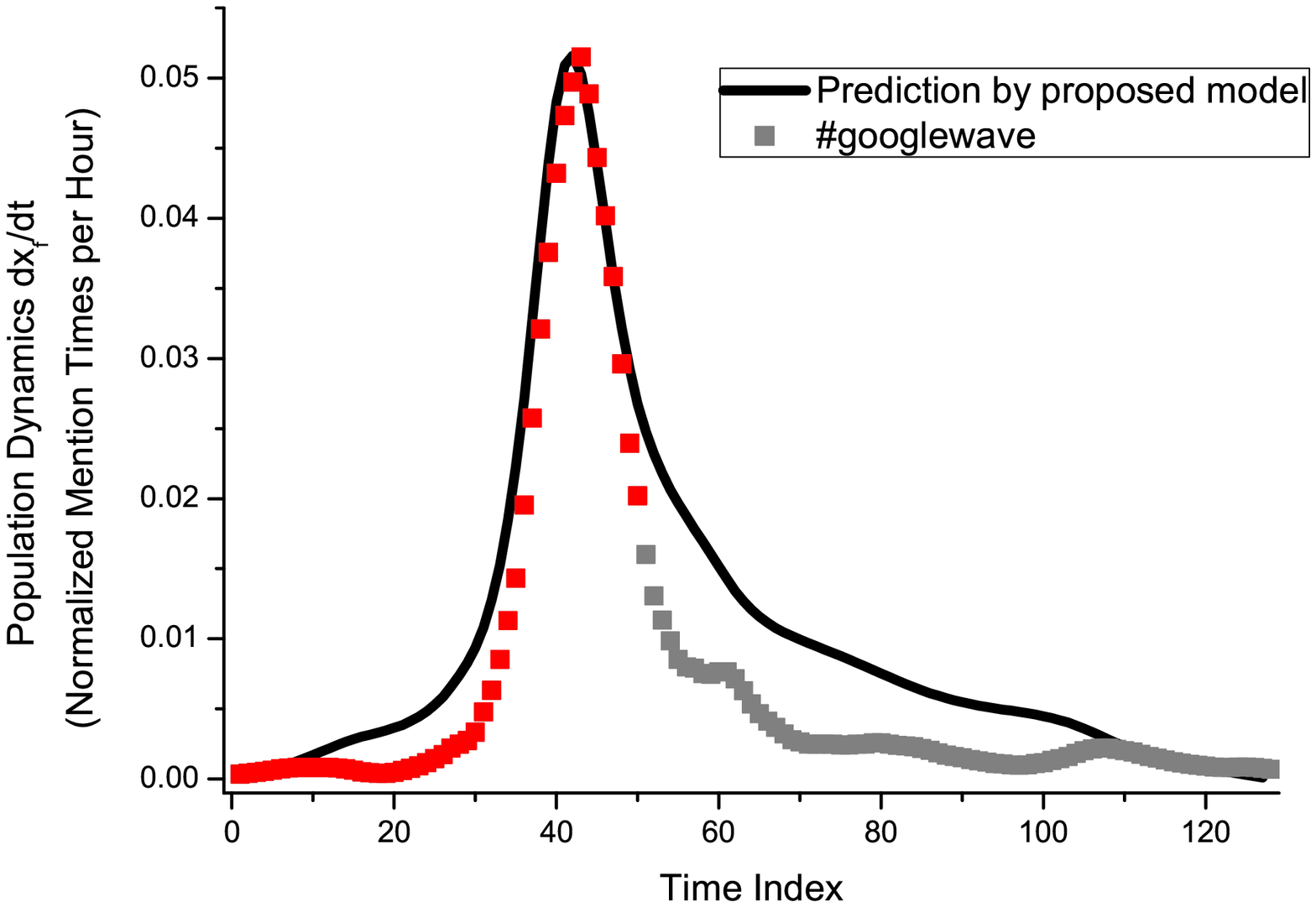,width=4.3cm}}
  \vspace{-0.1cm}
  \center{\scriptsize{(c) Curve fitting using 40\% data.}}
\end{minipage}
\hfill
\begin{minipage}[t]{0.24\linewidth}
  \centering
  \centerline{\epsfig{figure=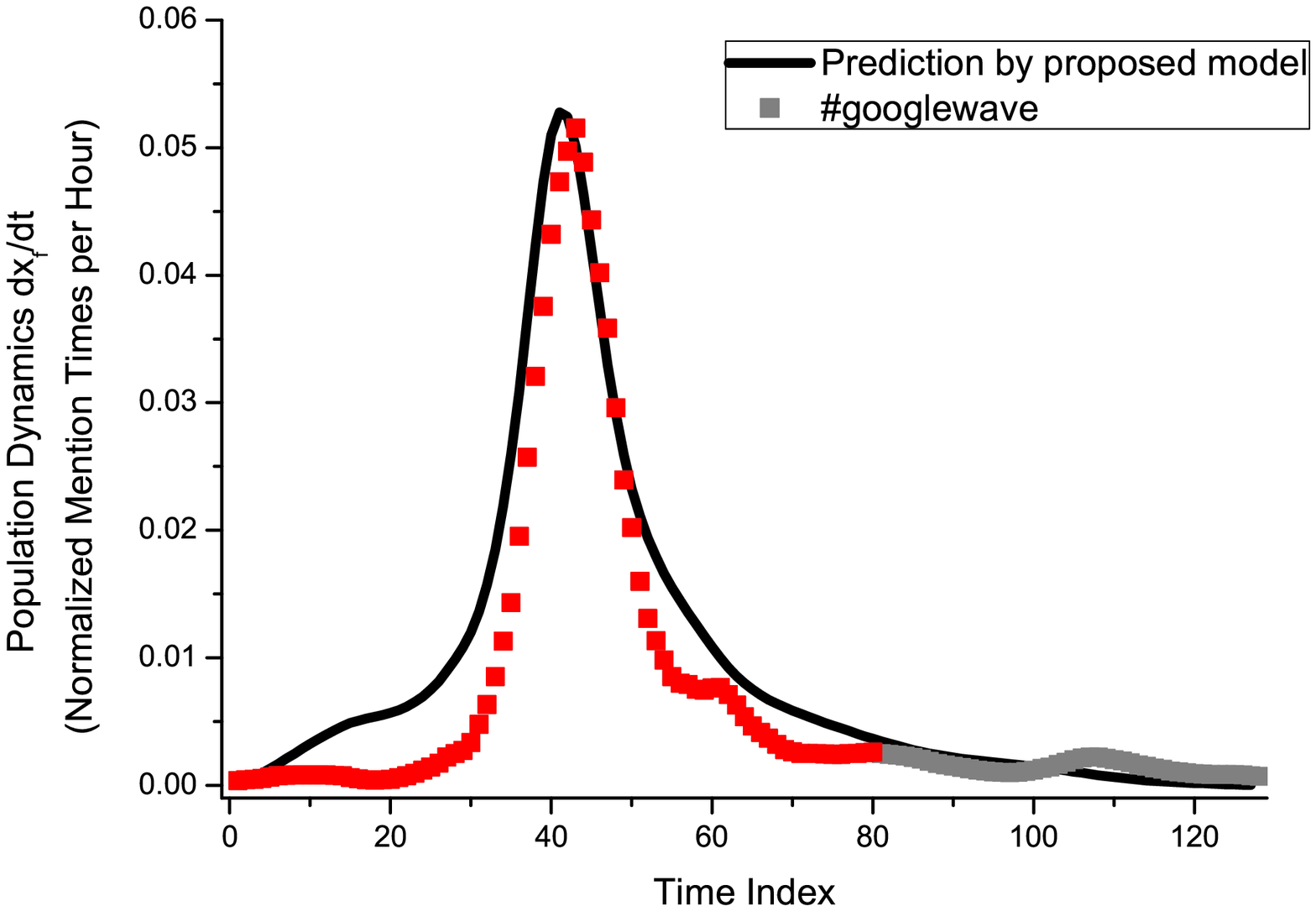,width=4.3cm}}
  \vspace{-0.1cm}
  \center{\scriptsize{(d) Curve fitting using 60\% data.}}
\end{minipage}
\caption{Prediction simulation restuls.}\label{predict}
\vspace{-5mm}
\end{figure*}

Finally, we conduct another experiment to verify that our proposed model can predict the dynamics of information diffusion. In this experiment, we only use parts of data to estimate the payoff matrix and check whether our model with the estimated payoff matrix can predict the remaining diffusion dynamics. In Fig.\,\ref{predict}-(a), we plot the prediction results for the hashtag \#googlewave, where only 25 percents of data are used for the payoff matrix estimation, i.e., the data denoted by solid red square. We can see that when only 25\% data are used, the prediction is effective until time index $40$, i.e., in the near future. Then, in Fig.\,\ref{predict}-(b) where 30 percents of data are used, the peak time can be precisely predicted and the prediction is effective until time index $45$. In Fig.\,\ref{predict}-(c) and (d) where 40 and 60 percents of data are used respectively, the predictions become more and more precise. This is because as long as the peak value and time index of the dynamics are included for payoff estimation, the precision of prediction would increase to a large extent since the peak value and time index are the most important information for the dynamics.

\section{Conclusion}

In this paper, we formulate the dynamics of information diffusion over social networks using evolutionary game theory. We defined the players, strategies and payoff matrix in this problem, and highlighted the correspondence between the EGT and information diffusion. Three kinds of networks, complete, uniform and non-uniform degree social networks, were analyzed with the derivation of information diffusion dynamics. Moreover, we also analyzed two representative networks: the Erd\H{o}s-R\'enyi random network and the Barab\'asi-Albert scale-free network. To validate our theoretical analysis, we conducted experiments on synthetic networks, real-world Facebook networks, as well as Twitter hashtags dataset. All the experiment results were consistent with corresponding theoretical results, which corroborated that our proposed EGT model is effective and practical for modeling the dynamics of information diffusion problem.


\appendices

\section{Proof of Theorem 2}\label{prooftheorem2}

The \emph{Case 1} and \emph{Case 2} listed in Section \ref{twocases} summarize two possible changes of the global network state $x_f$. For \emph{Case 1}, according to the BD strategy update rule, it is corresponding to the instance that when a user adopting strategy $\bm S_f$ is selected for reproduction and replaces a neighbor adopting strategy $\bm S_n$. Suppose the user has $k_f$ neighbors adopting strategy strategy $\bm S_f$ and $k-k_f$ neighbors adopting strategy strategy $\bm S_n$. In the BD strategy update rule, the global selection probability is proportional to the fitness. In such a case, the user adopting strategy $\bm S_f$ is selected with probability $x_f\Psi_f/\overline\Psi$ and the replacement probability is $(k-k_f)/k$, where $\Psi_f$ denotes the fitness of the user adopting strategy $\bm S_f$ and $\overline\Psi$ denotes the average fitness of the whole population. In such a case, the expected occurrence probability of \emph{Case 1} is
\begin{equation}
P^{\mbox{\scriptsize{BD}}}_{1}=\sum_{k_f=0}^k\frac{k!}{k_f!(k-k_f)!}x_{f|f}^{k_f}(1-x_{f|f})^{k-k_f}\frac{x_f\Psi_f}{\overline\Psi}\frac{k-k_f}{k},\label{case1}
\end{equation}
which is also the expected probability of the global network state $x_f$ increasing $1/N$. Similarly, for \emph{Case 2}, it is corresponding to the instance that when a user adopting strategy $\bm S_n$ is selected for reproduction and replaces a neighbor adopting strategy $\bm S_f$. As a dual expression of \ref{case1}, the expected occurrence probability of \emph{Case 2} is
\begin{equation}
P^{\mbox{\scriptsize{BD}}}_{2}=\sum_{k_f=0}^k\frac{k!}{k_f!(k-k_f)!}x_{f|n}^{k_f}(1-x_{f|n})^{k-k_f}\frac{(1-x_f)\Psi_n}{\overline\Psi}\frac{k_f}{k},\label{case2}
\end{equation}
which is also the expected probability of the global network state $x_f$ decreasing $1/N$. $\Psi_n$ denotes the fitness of the user adopting strategy $\bm S_n$.

We assume that there are $N$ unit period in each time slot, i.e., there are $N$ sub-slots and only one update occurs in each sub-slot. Combining (\ref{case1}) and (\ref{case2}), the expected changes of the global network state $x_{f}$ in one time slot is
\begin{align}
\dot x_{f}=&\ \sum_{k_f=0}^k\frac{k!}{k_f!(k-k_f)!}\Big\{x_{f|f}^{k_f}(1-x_{f|f})^{k-k_f}\frac{x_f\Psi_f}{\overline\Psi}\frac{k-k_f}{k}\nonumber\\
&\ -x_{f|n}^{k_f}(1-x_{f|n})^{k-k_f}\frac{(1-x_f)\Psi_n}{\overline\Psi}\frac{k_f}{k}\Big\},\label{29}
\end{align}
\begin{equation}
\mbox{where }\left\{
\begin{array}{l}
\Psi_f=1-\alpha+\alpha\big[k_fu_{ff}+(k-k_f)u_{fn}\big],\vspace{2mm}\\
\Psi_n= 1-\alpha+\alpha\big[k_fu_{fn}+(k-k_f)u_{nn}\big].
\end{array}\right.\label{30}
\end{equation}
By substituting (\ref{30}) into (\ref{29}), we can simplify the dynamics of global network state as
\begin{align}
\dot x_{f}=&\ \frac{\alpha x_f(1-x_{f|f})(k-1)}{\overline\Psi}\cdot\nonumber\\
&\ \big[(u_{ff}-u_{fn})x_{f|f}-(u_{nn}-u_{fn})(1-x_{f|n})\big].\label{31}
\end{align}
Notice that the numerator of (\ref{31}) is dominant by $\alpha$ which is sufficiently small compared with 1 in the weak selection scenario, while the denominator $\overline\Psi=1-\alpha+\alpha\overline U$ is dominant by $1$. In such a case, by substituting the local equilibria (\ref{xffs}) into (\ref{31}), we can approximate (\ref{31}) as
\begin{align}
\dot x_{f}\doteq&\ \alpha x_f(1-x_{f|f})(k-1)\cdot\nonumber\\
&\ \big[(u_{ff}-u_{fn})x_{f|f}-(u_{nn}-u_{fn})(1-x_{f|n})\big]\nonumber\\
= &\ \frac{\alpha (k-2)}{(k-1)}x_f(1-x_f)\big[(k-2)(u_{ff}-2u_{fn}+u_{nn})x_f\nonumber\\
&\ +u_{ff}+(k-2)u_{fn}-(k-1)u_{nn}\big].\label{bddynamic}
\end{align}

\section{Proof of Theorem 3}\label{prooftheorem4}

In this proof, we will first extend the analysis of information diffusion dynamics to that under the DB and IM strategy update rules, respectively. Then, we will show the equivalence by comparing the approximated expressions of diffusion dynamics under three kinds of rules, when the network degree is sufficiently large.

\subsection{Diffusion Dynamics under DB Rule}

Similar to the analysis in the BD strategy update rule, for the diffusion dynamics under DB rule, we also need to first calculate the stable point of the influence dynamics, i.e., the local equilibria of the information diffusion, and then calculate the population dynamics by analyzing the two cases listed in Section \ref{twocases}. Note that since the stable point of the influence dynamics under the DB strategy update rule is the same as that under the BD rule as shown in (\ref{xffs}), the detailed derivation is omitted here. Moreover, since the DB strategy update rule is dual with the BD rule, the population dynamics analysis under those two rules are also dual with each other. Recall that \emph{Case 1} in Section \ref{twocases} is corresponding to the instance that when a user adopting strategy $\bm S_n$ is selected for deviation and replaced by a neighbor adopting strategy $\bm S_f$. According to the DB strategy update rule, the expected occurrence probability of \emph{Case 1} is
\begin{equation}
P^{\mbox{\scriptsize{DB}}}_{1}=\sum_{k_f=0}^k\frac{k!}{k_f!(k-k_f)!}x_{f|n}^{k_f}(1-x_{f|n})^{k-k_f}\frac{(1-x_f)\Psi_f}{\overline\Psi_1}\frac{k_f}{k},\label{case11}
\end{equation}
Similarly, the \emph{Case 2} in Section \ref{twocases} is corresponding to the instance that when a user adopting strategy $\bm S_f$ is selected for deviation and replaced by a neighbor adopting strategy $\bm S_n$, the expected occurrence probability of which is
\begin{equation}
P^{\mbox{\scriptsize{DB}}}_{2}=\sum_{k_f=0}^k\frac{k!}{k_f!(k-k_f)!}x_{f|f}^{k_f}(1-x_{f|f})^{k-k_f}\frac{x_f\Psi_n}{\overline\Psi_2}\frac{k-k_f}{k}.\label{case21}
\end{equation}
Combining (\ref{case11}) and (\ref{case21}), the expected changes of the global network state $x_{f}$ in one time slot is
\begin{align}
\dot x_{f}=&\ \sum_{k_f=0}^k\frac{k!}{k_f!(k-k_f)!}\Big\{x_{f|n}^{k_f}(1-x_{f|n})^{k-k_f}\frac{(1-x_f)\Psi_f}{\overline\Psi}\frac{k_f}{k}\nonumber\\
&\ -x_{f|f}^{k_f}(1-x_{f|f})^{k-k_f}\frac{x_f\Psi_n}{\overline\Psi}\frac{k-k_f}{k}\Big\},\label{291}
\end{align}
where
\begin{equation}
\left\{
\begin{array}{l}
\!\!\!\Psi_f\!=\!1-\alpha+\alpha\big[(k-1)(x_{f|f}u_{ff}+(1-x_{f|f})u_{fn})+u_{fn}\big],\vspace{2mm}\\
\!\!\!\Psi_n\!=\!1-\alpha+\alpha\big[(k-1)(x_{f|n}u_{fn}+(1-x_{f|n})u_{nn})+u_{fn}\big].
\end{array}\right.\label{301}
\end{equation}
By substituting (\ref{301}) and the local equilibria (\ref{xffs}) into (\ref{291}) and adopting the similar approximation in Appendix A, we can simplify (\ref{291}) as
\begin{align}
\dot x^{\mbox{\scriptsize DB}}_{f}= &\ \frac{\alpha (k-2)(k+1)}{k(k-1)}x_f(1-x_f)\big[(k-2)(u_{ff}-2u_{fn}\nonumber\\
&\ +u_{nn})x_f+u_{ff}+(k-2)u_{fn}-(k-1)u_{nn}\big],\label{dbdynamic}
\end{align}
which is the diffusion dynamics under DB strategy update rule.

\subsection{Diffusion Dynamics under IM Rule}

The diffusion dynamics under IM strategy update rule can be derived in a similar way. First, for the stable point of the influence dynamics, it is the same as that under BD and DB strategy update rules as shown in (\ref{xffs}). Secondly, the analysis of population dynamics under IM strategy update rule is quite similar with that under the DB rule, since the only difference between the IM and DB strategy update rules is that the user may imitate his/her own strategy in the IM rule as shown in Fig.\,\ref{fig3}-(b) and (c). Thus, the diffusion dynamics under IM strategy update rule can be derived by taking into account the probability of one user remaining his/her strategy unchanged, which is
\begin{align}
\dot x^{\mbox{\scriptsize IM}}_{f}= &\ \frac{\alpha k(k-2)(k+3)}{(k-1)(k+1)^2}x_f(1-x_f)\big[(k-2)(u_{ff}-2u_{fn}\nonumber\\
&\ +u_{nn})x_f+u_{ff}+(k-2)u_{fn}-(k-1)u_{nn}\big].\label{imdynamic}
\end{align}

\subsection{Equivalence}

Comparing the diffusion dynamics under three kinds of strategy update rules, i.e., (\ref{bddynamic}), (\ref{dbdynamic}) and (\ref{imdynamic}), we can see that they have exactly the same form of expressions with the only difference in terms of coefficient, i.e., the BD rule is with coefficient $\frac{\alpha (k-2)}{(k-1)}$, the DB rule is with coefficient $\frac{\alpha (k-2)(k+1)}{k(k-1)}$ and the IM rule is with coefficient $\frac{\alpha k(k-2)(k+3)}{(k-1)(k+1)^2}$. In such a case, when the network degree $k$ is sufficiently large, all those three coefficients would tend to $\alpha$, which means that the diffusion dynamics over uniform degree networks under those three kinds of strategy update rules are equivalent with each other. This completes the proof of the theorem.

\section{Proof of Theorem 4}\label{prooftheorem3}

In the derivation of local dynamics of the uniform degree networks, i.e., the proof of \emph{Lemma 1}, the ``$k$'' in (\ref{increase}) and (\ref{decrease}) represents the degree of the replaced neighbor. In the non-uniform scenario, the degree of the replaced neighbor obeys some distribution, instead of constant $k$. Note that if a pair is selected by random, the degree distribution of the user on the specific pair is not $\lambda(k)$ but rather $\frac{k\lambda(k)}{\sum_{k=0}^{+\infty}k\lambda(k)}$ \cite{degreec}. In such a case, the average degree of the replaced neighbor is
\begin{equation}
\sum_{k=0}^{+\infty}k\frac{k\lambda(k)}{\sum_{k=0}^{+\infty}k\lambda(k)}=\frac{\overline {k^2}}{\overline k},\label{53}
\end{equation}
where $\overline {k^2}=\sum_{k=0}^{+\infty}k^2\lambda(k)$ is the expectation of $k^2$.
For the non-uniform scenario, when calculating the average increase and decrease of the $(\bm S_f,\bm S_f)$ pairs in a unit update period, we need take expectations with respect to the ``$k$'' in (\ref{increase}) and (\ref{decrease}). Moreover, the denominator in (\ref{denom}), which represents the total number of edges in the network, should be $N\overline k/2$. Thus, the local dynamics of the non-uniform degree networks becomes
\begin{align}
\dot x_{f|n}\!=&\frac{2}{\overline k}\Big\{\Big(\frac{\overline {k^2}}{\overline k}-1\Big)\big[x_{f|f}x_{f|n}+x_{f|n}(1-x_{f|n})\big]\!-\!\frac{\overline {k^2}}{\overline k}x_{f|n}\Big\},\\
\dot x_{f|f}\!=&\frac{2}{\overline k}\Big\{1\!+\!\Big(\frac{\overline {k^2}}{\overline k}\!-\!1\Big)\big[x_{f|f}x_{f|f}\!+\!x_{f|n}(1\!-\!x_{f|f})\big]\!-\!\frac{\overline {k^2}}{\overline k}x_{f|f}\Big\}.
\end{align}
The local equilibrium of the information diffusion dynamics over non-uniform degree networks becomes
\begin{equation}
x^*_{f|n}=\frac{(\overline {k^2}-2\overline k)x_f}{{\overline {k^2}}-\overline k},\quad
x^*_{f|f}=\frac{({\overline {k^2}}-2\overline k)x_f+\overline k}{\overline {k^2}-\overline k}.\label{xffs2}
\end{equation}

For the global dynamics, we also need to consider the distribution of users' degrees. Since the ``$k$'' in (\ref{31}) represents the degree of the selected user, we can directly take expectation on it and obtain the global dynamics of the non-uniform degree networks as
\begin{align}
\dot x_{f}=&\ \frac{\alpha x_f(1-x_{f|f})(\overline k-1)}{\overline\Psi}\cdot\nonumber\\
&\ \big[(u_{ff}-u_{fn})x_{f|f}-(u_{nn}-u_{fn})(1-x_{f|n})\big].\label{100}
\end{align}
Similarly, by substituting (\ref{xffs2}) into (\ref{100}) and approximating $\overline \Psi$, we have
\begin{align}
\dot x_{f}= &\ \frac{\alpha ({\overline k}-1)({\overline {k^2}}-2{\overline k})}{({\overline {k^2}}-{\overline k})^2}x_f(1-x_f)\big[({\overline {k^2}}-2{\overline k})(u_{ff}-2u_{fn}\nonumber\\
&\!\!\!\! +u_{nn})x_f+{\overline k}u_{ff}+({\overline {k^2}}-2{\overline k})u_{fn}-({\overline {k^2}}-{\overline k})u_{nn}\big].\!\!\!\!
\end{align}

\bibliographystyle{IEEEtran}

\bibliography{list}

\end{document}